\documentclass[12pt,preprint]{aastex}
\usepackage{natbib}
\usepackage{epsfig}
\begin{document}

\title{The Bulge Radial Velocity Assay ({\it BRAVA}): I. Sample Selection and a Rotation Curve}
\author{ Christian D. Howard\altaffilmark{1,2},
R. Michael Rich\altaffilmark{1},
David B. Reitzel\altaffilmark{1,2},
Andreas Koch\altaffilmark{1},
Roberto De Propris\altaffilmark{3},
HongSheng Zhao\altaffilmark{4}}

\altaffiltext{1}{Department of Physics and Astronomy, University of California, Los Angeles, CA 90095-1562}
\altaffiltext{2}{Visiting Astronomers, Cerro Tololo Inter-American Observatory.  CTIO is operated by AURA, Inc.\ under contract to the National Science Foundation.}
\altaffiltext{3}{Cerro Tololo Inter-American Observatory, Casilla 603, La Serena, Chile}
\altaffiltext{4}{SUPA, School of Physics and Astronomy, University of St. Andrews, KY16 9SS, UK}

\begin{abstract}
Results from the ongoing Bulge Radial Velocity Assay ({\it BRAVA}) are presented.  {\it BRAVA} uses M red giant stars, selected from the {\it 2MASS} catalog to lie within a bound of reddening corrected color and luminosity, as targets for the Cerro Tololo Inter-American Observatory 4-m Hydra multi-object spectrograph.  Three years of observations investigate the kinematics of the Galactic bulge major ($-$10$^\circ$$<${\it l}$<$$+$10$^\circ$, {\it b}$=$$-$4$^\circ$) and minor ($-$6$^\circ$$<${\it b}$<$$+$5$^\circ$, $-$0.4$^\circ$$<${\it l}$<$$-$0$^\circ$) axes with $\sim$3300 radial velocities from 32 bulge fields and one disk field.  We construct a longitude-velocity plot for the bulge stars and find that, contrary to previous studies, the bulge does not rotate as a solid body; from $-$4$^\circ$$<${\it l}$<$$+$4$^\circ$ the rotation curve has a slope of roughly 100 km~s$^{-1}$~kpc$^{-1}$ and flattens considerably at greater {\it l}, reaching a maximum rotation of 75 km~s$^{-1}$.  We compare our rotation curve and velocity dispersion profile to both the self-consistent model of Zhao (1996) and to N-body models; neither fits both our observed rotation curve and velocity dispersion profile.  We place the bulge on the plot of (V$_{max}$/$\sigma$) vs. $\epsilon$ and find that the bulge lies near the oblate rotator line, and very close to the parameters of NGC 4565, an edge-on spiral galaxy with a bulge similar to that of the Milky Way.

We find that our summed velocity distribution of bulge stars appears to be sampled from a Gaussian distribution, with $\sigma=116 \pm 2$ km~s$^{-1}$ for our summed bulge fields.   Furthermore, the high precision of our radial velocities ($\sim$5 km~s$^{-1}$) allows us to investigate hints of cold kinematic features that were seen in a number of the line-of-sight velocity distributions from our earlier observations.  In the past, optical radial velocity studies in the bulge have not emphasized high precision, because of the large velocity dispersion and the expectation that the short orbital periods would erase any cold structures in well under a gigayear. Thus our precision is sufficient to enable a search for more cold streams analogous to those associated with the Sagittarius dwarf spheroidal galaxy; some candidate cold features were seen and follow-up observations are reported herein.

\end{abstract}
\section{Introduction}

\begin{figure*}[t]
\plotone{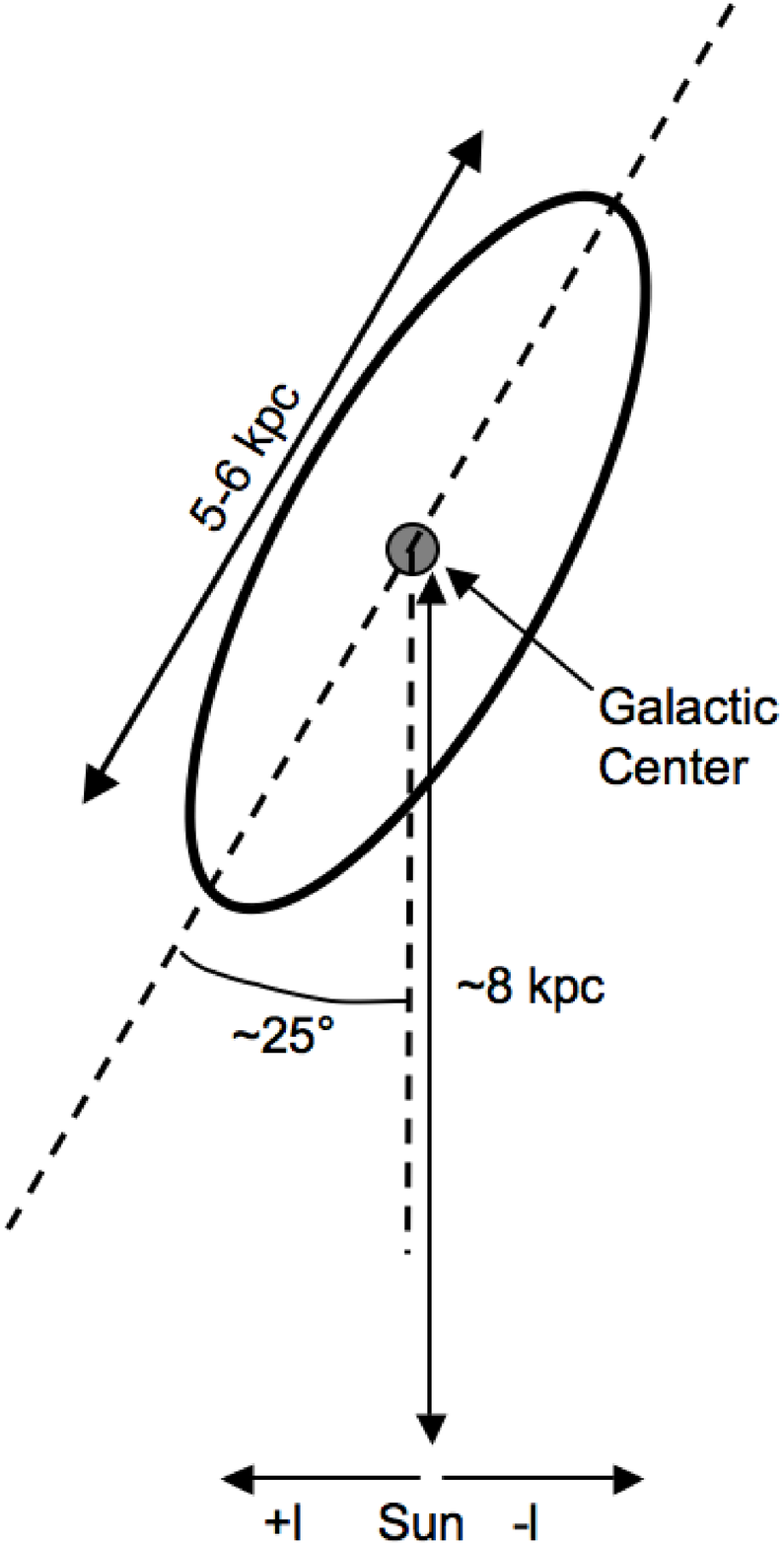}
\caption{A simple schematic of the bulge/bar orientation with respect to the Sun/Galactic center line of sight. Exact values of the angle are uncertain
but range from $\sim$15$^\circ$ to $\sim$35$^\circ$ (Alcock et al. 2000;Gerhard et al. 2002). On this diagram, positive Galactic latitude ({\it b}) points up out of the page.}
\label{schematic}
\end{figure*}

The status of the Milky Way bulge as a distinct stellar population had historically been well established by Baade's surveys of RR Lyrae stars, and the well known concentration of red giants toward the Sagittarius region (Blanco 1965).  Furthermore, the image of the Galaxy produced by the {\sl COBE} satellite (Weiland et al.1984; Dwek et al. 1995;  Arendt et al. 1998) depicted an unambiguous, asymmetric, peanut-shaped bulge.  Blitz \& Spergel (1991) modeled the $2.4\mu \rm m$ balloon borne data of Matsumoto et al. (1982) as a bar which shows the hallmark thickening of the bulge at positive Galactic longitude that is seen in later studies (see e.g. Launhardt et al. 2002).  While the discovery of a bar in the stellar distribution was somewhat of a surprise, gas motions in the inner 2 kpc have long been modeled as due to the potential of tilted bar  (e.g. Liszt \& Burton 1980).  Optical surveys find roughly 1/3 of spirals are barred (Sellwood \& Wilkinson 1993); this fraction rises to 60\% or more in the infrared (Menendez-Delmestre et al. 2007), so it is perhaps not surprising that one should find evidence of bar structure in the Milky Way.  In fact, the evidence for a triaxial (barred) bulge is convincing.  The COBE 2$\mu$m light distribution is modeled as a bar oriented toward positive Galactic longitude (Dwek et al. 1995, Binney et al. 1997).  Star counts of  red clump stars (Stanek et al 1997; Babusiaux \& Gilmore 2005) show a barlike structure, while microlensing events further show a central bar that is pointed roughly in the direction of the Sun (Figure \ref{schematic}).  

There is also the issue of bar survival; Sheth et al. (2007) find that the bar fraction declines dramatically with redshift to half the Hubble time, with only the most massive galaxies hosting bars by $z\sim 0.8$.  This fact is to be contrasted with growing evidence that the bulge is old from a chemical (Rich 1990; McWilliam \& Rich 1994; Matteucci et al. 1999; Fulbright et al. 2007; Lecureur et al. 2007) and stellar populations (Ortolani et al. 1995; Kuijken \& Rich 2002; Zoccali et al. 2003) perspective.   There is also continuing debate over whether the Milky Way bulge/bar system is best categorized as a pseudobulge (Kormendy \& Kennicutt 2004).  As evidence of a bar in the Milky Way has grown (see e.g. Gerhard et al. 2002) it has become clear that the Galactic bulge offers a potential laboratory to investigate the dynamics of the nearest bar in detail with radial velocities and eventually with proper motions as well as abundances.  This recent evidence has prompted renewed interest in the dynamical modeling of the Galactic bar/bulge system in order to determine, in part, the dynamical formation and evolution of galaxies e.g., the unsolved issue of whether a Cold Dark Matter halo cusp is compatible with the microlensing and rotation of the Milky Way bar (Zhao, Spergel, Rich 1995, Klypin, Zhao, Somerville 2002).

However, Zhao's (1996) dynamical model paper notes the lack of comprehensive surveys as a problem in comparing the model against data; until recently there simply has not been enough data to properly constrain the few dynamical models of the bulge available.  Indeed, many efforts have modeled the 2 micron light in projection (cf. Gerhard 2002), but surveys of the dynamics have generally either emphasized very rare evolved stars (e.g. SiO masers; Izumiura et al. 1995; OH/IR or IRAS stars; Menzies 1995) or use K/M giants. Previous studies are restricted to small isolated fields due to the problems of crowding, large and variable extinction, lack of wide field surveys in the red, and the presence of a contaminating disk population.  Minniti (1992) used K giants, but at the time, uniform selection of candidates was an issue, and concerns were raised about disk contamination, especially for the field at ({\it l,b})=(12$^\circ$,3$^\circ$).  Consequently, past surveys have had insufficient statistics to properly probe the dynamics of fields at Galactic latitude $|$ {\it l} $|$ $<$ 5$^\circ$ (see Table 3 of Zhao 1996 for detailed summary) in order to provide a constraint for dynamical models.

Partly in response to this problem, Beaulieau et al. (2000) undertook the first dynamical survey of the bulge aimed at isolating a uniform sample of kinematic probes while at the same time covering a wide area.  Their study combines PNe from a new survey and those found in existing catalogs to span a wide range in Galactic longitude and to study both the Northern and Southern portions of the bulge.  They argue that the easily observed PNe have no metallicity bias.  However, PNe are short-lived and are therefore rare, and their population membership (and especially distances) are uncertain.  Rich et al. (2007) noted (and we confirm) that the bulge PNe have lower velocity dispersion than other populations, consistent with a partial population of non-bulge members.

The dynamical model for the bulge/bar has a number of important implications. Large samples of uniform radial velocity data are still of great value in constraining the bar versus axisymmetric models, and the nature of the orbit families supporting the bar. Further, the interpretation of the microlensing events in the bulge depends on the use of an accurate dynamical model (Han \& Gould 2003). The recent discovery of planetary transit host stars in the bulge (Sahu et al. 2006) gives an additional incentive to improve our knowledge of the bulge/bar model, as it is debated whether stars on non-circular orbits are incapable of bearing planets, if their orbits stray too near the Galactic center.  The relatioinship to the disk, thick disk, and halo populations is of great interest.  Only with a realistic dynamical model can the issue of long term stability be explored.  There is also the issue of whether the bar formed due to the buckling of the disk (see e.g. Raha et al. 1991) or is a yet more ancient population.  

The availability of bulge photometry and high precision astrometry from the Two-Micron All-sky Survey (2MASS; Skrutskie et al. 2006) along with the Schlegel et al. (1998) all sky reddening map offered, for the first time, the possibility of easily selecting a large sample of kinematic probes from the very population responsible for the 2$\mu$m radiation comprising the projected bulge/bar (Dwek et al. 1995; Launhardt et al. 2003; Figs. 1,2).   The {\it Bulge Radial Velocity Assay} or {\it BRAVA} (Rich et al. 2007) was conceived as a survey of the line of sight velocity distribution of red giants across the bulge, to be compared with self-consistent dynamical models like that of Zhao (1996).  {\it We emphasize that while we report the first and second moments of the BRAVA fields in this paper, the ultimate aim is to study the Line of Sight Velocity Distribution (LOSVD) and to compare  it with dynamical models.}   A further additional aim of {\it BRAVA} is to explore correlations between kinematics and abundances.  Note that there is already a hint from one proper motion field study (Soto, Rich, \& Kuijken 2007) that predominantly metal rich stars are observed to show a vertex deviation in their velocity ellipsoid and may therefore be predominantly responsible for the support of the bar.  If $\rm [\alpha/Fe] $ can be derived in addition to
[Fe/H], it may be possible to use the $\alpha-$enhancement as a secondary criterion to distinguish bulge from thick disk stars, via chemistry (see e.g. Fulbright et al. 2007).

Here we report results from a survey based on red giants, which comprise the bulk of the 2.4$\mu \rm m$ light of the bulge.  M giants offer a dynamical probe that is populous, luminous, traces the light, and is easily utilized for velocity measurements.  M giants are also extremely well studied (Frogel \& Whitford 1987) and have been identified, and their giant branch characterized, over the whole of the bulge (Frogel et al. 1990), from $-3^\circ$ to $-12^\circ$ (420 to 1675 pc, adopting $R_0=8$ kpc).   The issue of the abundance "bias" raised by Beaulieau et al. is essentially settled based on high dispersion infrared spectra (see Rich \& Origlia 2005, and Rich, Origlia, \& Valenti 2007).  M giants follow roughly the abundance distribution of the K giants; given their significant contribution to the integrated light, we are confident that these stars are excellent dynamical probes.   The M giants, being luminous, are also usable as probes even in regions of high extinction, and for exploring the fringes of the bulge, as Frogel et al. (1990) as well as the 2MASS catalog, illustrate.

Early results from our survey are given in Rich et al. (2007a,b).  Here, we report on the first three-year results from {\sl BRAVA} and also describe the sample selection and measurement uncertainties in detail.   At this time, we present $\sim$3300 spectra for which we have obtained a radial velocity precision of $\sim$5 km~s$^{-1}$ for our most recent data.   Future papers in this series will address the kinematics of the bulge at $-8^\circ$ and a detailed comparison with the Zhao (1996) model (Howard et al. 2008 in prep.).  The long term plan is to make the entire survey, including the spectra, available on a public website.

In the following section we discuss our sample selection.  Section 3 discusses our observations, while Section 4 discusses our data reduction and velocity calibration.  Section 5 considers the fields and the issue of reddening.  Section 6 presents our results, while Section 7 gives our conclusions.

\section{Sample Selection}

The first use of M giants to undertake a rapid survey of the velocity dispersion in Baade's Window (BW) was that of Mould (1983) who used the du Pont 2.5 m telescope at Las Campanas, and the photon-counting Shectograph, to obtain spectra of 50 stars and measure a velocity dispersion of $\sigma=113\pm 11$ km sec$^{-1}$.  The approach used in our present work was pioneered by Sharples, Walker, \& Cropper  (1990, hereafter SWC), who used the AAT and a multifiber spectrograph to obtain spectra of a sample of 239 late-type M giants in BW.  SWC found these M giants to be a kinematically distinct group in the bulge, and consequently excellent probes for studying the dynamics of the Milky Way bulge.  Beaulieau et al. (2000) use the planetary nebulae (PNe) population of the bulge as dynamical tracers, using a sample of nearly 400 PNe in the extended region $-$30{$^\circ$} $<$ {\it l} $<$ 30{$^\circ$} and 3.3{$^\circ$} $<$ $|${\it b}$|$ $<$ 15{$^\circ$} to investigate the kinematics.  However, the brief lifetimes of PNe, uncertain population membership, and distance uncertainties make them a potentially problematic population to probe bulge kinematics.  As a result, dynamical models of the bulge/bar system are, to date, relatively weakly constrained by radial velocity data.  A better solution is to conduct a survey using red giant stars, following the SWC strategy, which are the most common type of luminous evolved star.  We choose to work brighter than the red clump for two reasons.  First, brighter giants can be studied in obscured fields, and in highly obscured fields, they can be studied in the infrared.  Second, clump stars lying closer than the bulge tend to be both bluer and brighter, due to the proximity as well as the lower reddening.  The RGB slants redward, the red edge of the RGB is comprised of the most metal rich red giants, largely uncontaminated by any other population.  

The most luminous red giants are cool enough to form titanium-oxide (TiO) molecules in their atmospheres, which is a signature of their spectral classification M.  Because M giants are extremely cool, their spectra are dominated by these TiO absorption lines, particularly towards the red end of the spectrum at 7055\AA, although other strong bandheads lie redwards of 8000\AA.  Additionally, other spectral features such as the prominent calcium triplet at 8498\AA, 8542\AA, and 8662\AA~make M giants good candidates for radial velocity probes and allow for subsequent metallicity measurements.  M giants, classified from low dispersion spectra,  are known to be ubiquitous throughout the bulge (Blanco, McCarthy, \& Blanco 1984; Frogel et al. 1990), and although faint at optical wavelengths due to the heavy extinction towards the Galactic center, TiO bands, and cool temperatures, they are relatively bright in the {\it I} band ($\sim$6500\AA~to $\sim$10000\AA).  {\it 2MASS} gives us adequate astrometry to use the M giant population as dynamical probes of the Galactic bulge; these stars' relatively short lifetimes makes their numbers $\sim $ 10 times less numerous than stars on the red giant branch and the clump, and consequently source confusion becomes a non-issue.  M giants are also more luminous; even in the $\sim 8000$\AA\ $I$ band where TiO bands are present, the M giant population is $\sim$ 5 mag brighter than the red clump, another factor that suppresses crowding and complications compared to the red clump giants.  This can easily be seen by comparing a 2MASS H-band image of Baade's Window with any optical image of that region.

When choosing stars as dynamical probes, one must be careful to minimize foreground contamination by main sequence disk stars and giants in the near disk, as well as to minimize any metallicity selection bias.  SWC observed Baade's Window and found that M giants with {\it I} $<$ 11.8 have a lower velocity dispersion and thus were likely to be disk members, whereas the stars with {\it I} $>$ 11.8 are likely bulge members, based on their kinematics.  When the survey fields are examined in {\it 2MASS}, the {\it K} vs. {\it J-K} color magnitude diagram (CMD) shows a clearly defined red giant branch (Figure \ref{cmd}).  Due to differential reddening towards the Galactic Bulge, the magnitude limit of {\it I} $<$ 11.8 corresponds roughly to {\it K} $<$ 8.2.  Therefore, we adopt a range of 9.25$>$ {\it K} $>$8.2 as the criterion for selecting {\it 2MASS} M giants; in principle, one can go even fainter, but the abundance of sources in the magnitude selection range is more than adequate, allows for less integration time, and avoids the red clump, an evolutionary path available only to stars with [Fe/H]$\geq -1$.  
 
\begin{figure*}[t]
\plotone{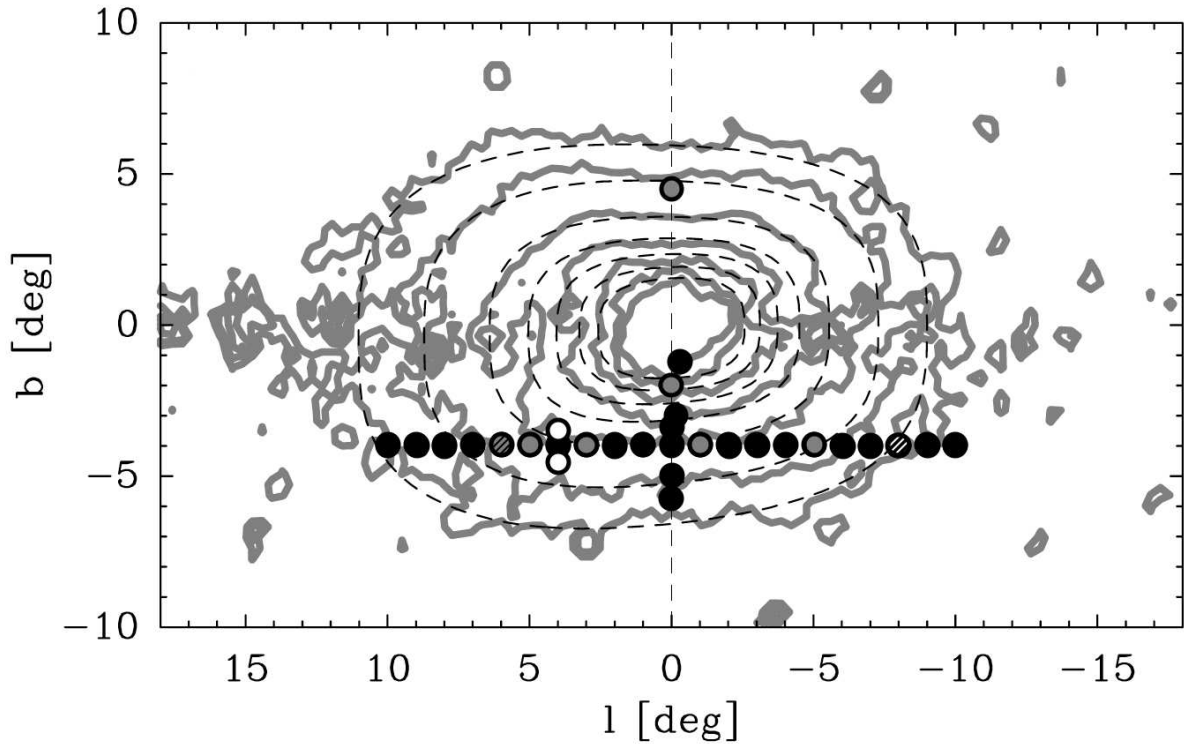}
\caption{Observed BRAVA fields,  up to April 2007, overplotted on the {\it COBE} 2$\mu$m image (Launhardt et al. 2002).  Grey circles represent fields observed in 2005, black circles are fields observed in 2006, and white circles are fields observed in early 2007.  The size of the circles corresponds to the 40$^{\prime\prime}$ field of view of the instrument.  Dashed circles represent fields observed over multiple years.  1{$^\circ$} corresponds to 140 pc for a distance of 8 kpc; the {\it b}$=-$4{$^\circ$} strip lies roughly 550 pc south of the Galactic center.  The Galactic plane is avoided due to heavy extinction.}
\label{survey}
\end{figure*}

\begin{figure*}[t]
\plotone{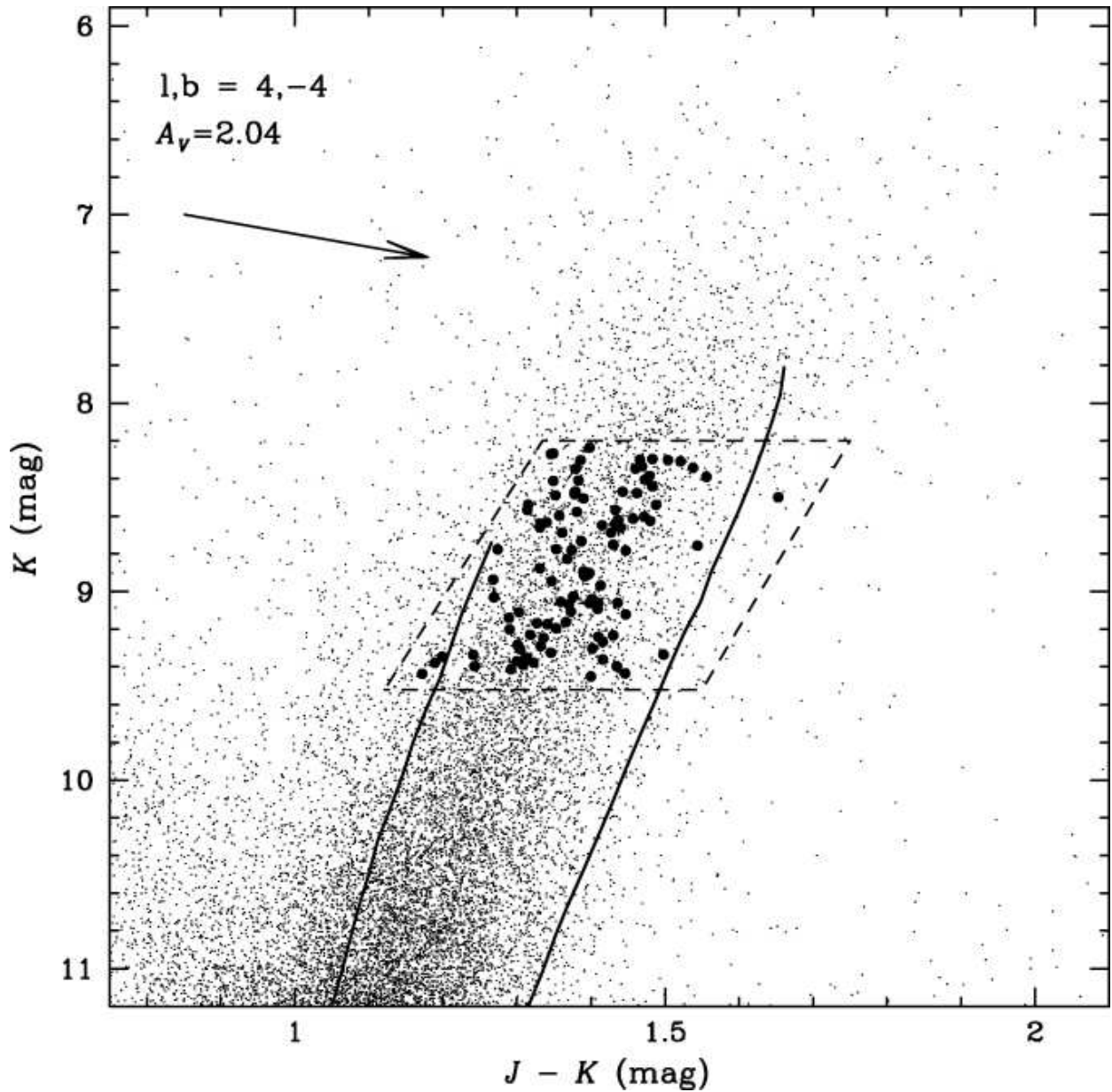}
\caption{Color-Magnitude Diagram of 2MASS targets in the field at (l,b)=(4{$^\circ$},$-$4{$^\circ$}).  Filled circles indicate stars for which we have spectra. Included are reddened isochrones (Girardi et al. 2002) for [Fe/H]$=-1.3$ and $-0.5$.  The reddening vector corresponds to E(J-K)=0.33 from the Schlegel et al. (1998) map.  The parallelogram indicates our selection region;  the blue cutoff rejects many objects that are closer than the bulge, which have lower reddening and are brighter than the red giant branch.} 
\label{cmd}
\end{figure*}

Because the reddening varies greatly in the galactic bulge (1.5 $<$ A$_{v}$ $<$ 5), a parallelogram-shaped region within the CMD is adjusted by eye in both magnitude and color to center on the expected locus of the distance and average reddening of each particular field, for which we used the extinction law calculator on the NED website (see Schlegel et al. 1998 for details). In fields with large extinction, the width of the parallelogram is widened to account for the large differential reddening.  This initial selection method assures bulge membership and minimizes any metallicity bias.  Our survey, to date, dynamically samples the Galactic bulge by observing red giants in fields at 1{$^\circ$} intervals along the major axis at {\it b} $=$$-$4{$^\circ$} along with the minor axis at {\it l}$\approx$0{$^\circ$} (Figure \ref{survey}).  

\subsection{Disk Contamination}
The major issue in the selection of bulge samples is the purity of a bulge sample.  The sightline crosses through the foreground disk, and these populations (mostly $<$ 1 Gyr).  Early samples were optically selected, and in those cases,
there may be considerable uncertainty in assigning population membership.  Sample contamination can arise from a number of sources.  Thick disk stars with similar abundances, kinematics, age, and distance, are indistinguishable
from the bulge.  Such a population must be included in models of the total luminosity and dynamics.  Thin disk populations are separable; they are younger and therefore, bluer.  Most such contamination arises blueward of the most metal poor isochrone.  But in highly reddened fields, differential reddening can scatter some thin disk stars into the
selection zone of the bulge; dereddening has removed most of these.  Foreground stars may overlay the selection region, by being closer and less reddened than the sample; this is a concern for the red clump, which is populous.  We are 3 mag brighter than the red clump, so this is not an issue.  The bar does have spatial depth, but all bar members will ultimately be modeled; kinematics as a function of photometric distance is beyond the scope of this paper.

We have empirically controlled for a possible contaminating population by determining how purely Gaussian the velocity distributions are, and we have also compared the kinematics of subdivisions of the population by bright and faint, and blue and red (Section 6).  The $l-v$ plot discussed in section 6 also shows no hints of extra subpopulations that might appear as linear features (a cold disk population).  We have also studied a disk field at ({\it l,b})$=(-30^\circ, -4^\circ)$, finding its velocity dispersion a factor of 2 lower than any in our sample (Section 5).  We are satisfied that our selection method yields a sample that is dominated by bulge/bar members and minimizes any disk contamination.  Future extension of a study like this to include measurements of [Fe/H] and composition might be able to bring additional tools to bear in the assignment of population membership.

\section{Observations}
\begin{figure*}[t]
\epsscale{0.8}
\plotone{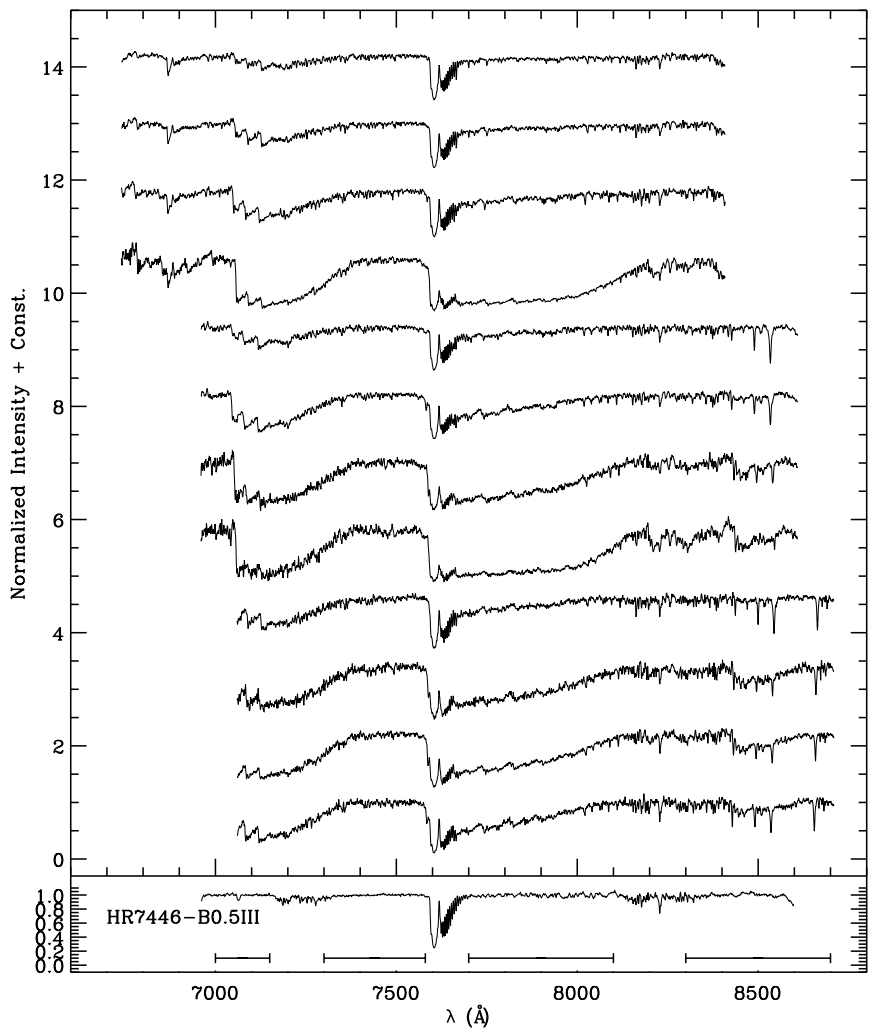}
\caption{A sample of the spectra observed from each year of observations, 2005 at top, moving down to 2007 at bottom.  The lower panel is a blue star observed in 2006 to illustrate where there are areas of significant atmospheric contamination.  The hash marks below the blue star mark regions used for cross correlation. Evident in the stellar spectra are the TiO bands at $\sim$7050\AA~, as well as the Telluric A band at $\sim$7600\AA.  In the 2006/2007 spectra the CaII triplet is seen at $\sim$8498\AA,$\sim$8542\AA, and, beyond the spectral range covered in 2006 but observed in 2007, the third line at $\sim$8663\AA.}
\label{sample_spec}
\end{figure*}

Observations began in 2005 and continued through 2007 (see Table \ref{log}), yielding 32 individual bulge fields and one disk field at ({\it l,b})$~$($-$30$^\circ$,$-$4$^\circ$).  We use the Hydra multifiber bench spectrograph at the Cassegrain focus of the Cerro Tololo Inter-American Observatory (CTIO) Blanco 4-m telescope.  On average, 103 stars are observed per field for a total of $\sim$3400 stars observed.  To take advantage of the red colors of M giants, we employ the KPGLD grating, blazed at 8500\AA~giving a dispersion 0.45\AA~per pixel with 2 pixel on chip binning, yielding an effective dispersion of 0.88\AA~per pixel and a full spectral range of $\sim$1800\AA.  In 2005, our central wavelength was $\sim$7600\AA~with an effective resolution of R$\sim$2800.  However, in 2006 we adjusted redward with a central wavelength of $\sim$7800\AA~in order to observe the calcium triplet missed in 2005.  In order to retain the TiO band at $\sim$7050\AA, we were only able to observe the first two lines of the triplet.  Despite this, very strong cross correlation peaks were obtained from the addition of the first two lines of the calcium triplet observed in 2006.  In 2007 we adjusted further redward, with a central wavelength of $\sim$7900\AA. Additionally, in 2006/2007 we utilized the 200$\mu$m slit plate giving us an increase in resolution (R$\sim$4200).  The light lost from the slit plate was determined to have little effect on our S/N due to the brightness of the sources.  Typically, we obtained 3 exposures of each field at 600 seconds each, although a few fields required 3x900 second exposures, or longer, due to poor observing conditions.  Each field has, on average, 105 successfully exposed M giants and $\sim$20 dedicated fibers that are used to obtain a sky background spectrum. Table \ref{log1} shows a summary of our observations including the number of reliable velocities measured in each field.  A sample selection of our stars from each year is shown in Figure \ref{sample_spec}, along with a blue star to show regions of telluric absorption.  The 2005 observations did not include any velocity standards, although we obtained 4 in 2006, and 2 in 2007 (Table \ref{standards}).

\section{Data Reduction and Velocity Calibration}
\begin{figure*}[t]
\plotone{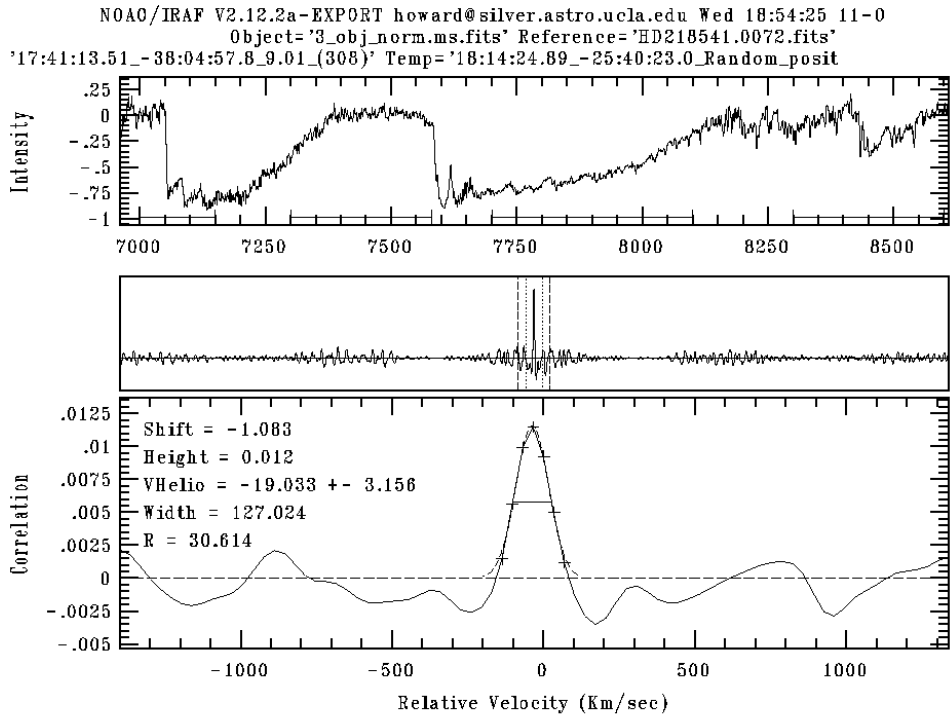}
\caption{A typical cross correlation peak as reported by the IRAF cross-correlation routine `fxcor`.  As can be seen in this example from 2006, the correlation peaks are quite strong, and yield Tonry-Davis R (TDR) parameters of ~10-30, yielding velocity errors between 5-15 km~s$^{-1}$, depending on the instrument set-up used.}
\label{fxcor}
\end{figure*}

The raw frames are first trimmed and overscan corrected using the IRAF routine `ccdproc`, and are bias corrected.  Flat fielding, sky subtraction, throughput correction, scattered light subtraction, and wavelength calibration are all accomplished through the IRAF task `dohydra'.  The spectra are binned to $\sim$34.5 km~s$^{-1}$~pixel$^{-1}$, co-added, and continuum normalized by a 2nd or 3rd order polynomial.  Radial velocities are then measured using the IRAF cross correlation routine `fxcor' utilizing a fourier filtering set to exclude both features exceeding
50 pixels (overall continuum) or smaller than 3 pixels (the spectral resolution).  Regions of the spectrum contaminated by telluric features (such as the atmospheric A band at $\sim$7600\AA) were omitted from the fitting, leaving roughly 60\% of our total spectrum usable for cross correlation; velocities are then determined by measuring a Gaussian fit to the strongest correlation peak (see fig \ref{fxcor}), and corrected to the heliocentric rest frame.

All 4 velocity standards observed in 2006 are used in the 2006 cross-correlations.   For the 22 2006 fields observed, 3 of the standards (HD 207076, HD 218541, and HD 203638) return individual stellar velocities that agree to better than 2 km~s$^{-1}$, on average, whereas the typical errors in stellar velocities reported by `fxcor', as determined by the Tonry-Davis R-value, are of order of 5-10 km~s$^{-1}$.  HD 177017 returns velocities that are offset from the aforementioned 3 standards by $\sim$7 km~s$^{-1}$ on average with a standard deviation of 1 km~s$^{-1}$.  Because  velocity offsets found with HD 177017 were constant from field to field in the 22 fields observed in 2006, final 2006 velocities were calculated by applying a zero-point shift of the velocities returned by HD 177017, and then taking the weighted average of all four standards.

\begin{figure*}[t]
\plotone{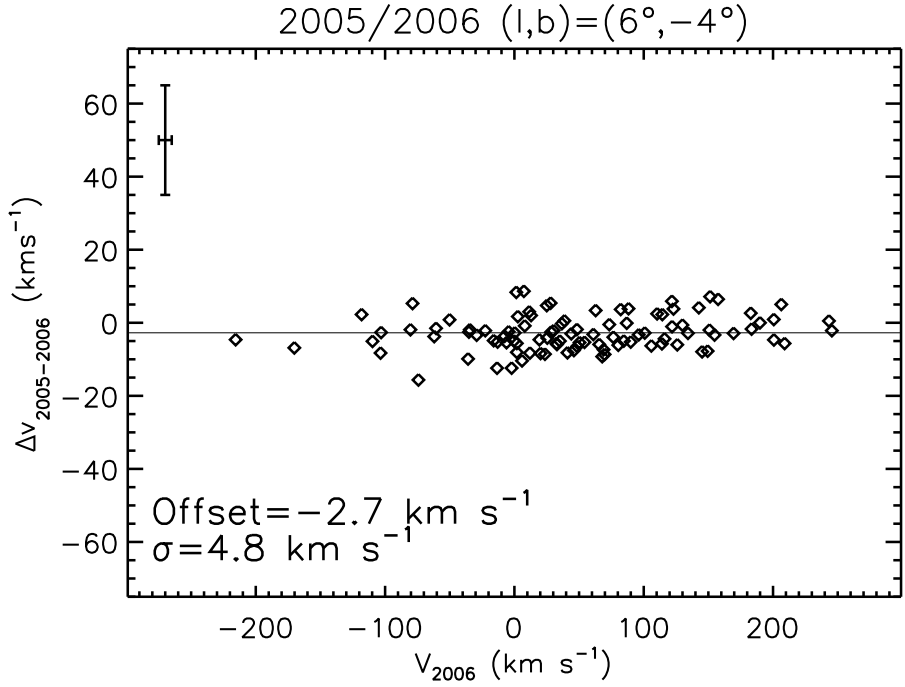}
\caption{Stellar radial velocity offset between 2005 and 2006 observations for field (l,b)=(6{$^\circ$},-4{$^\circ$}).  Our 2005/2006 individual stellar velocities show agreement to within 5km~s$^{-1}$.  We therefore adopt 5 km~s$^{-1}$ as our individdual stellar velocity error.  Since the dispersion we see in our fields is over an order of magnitude larger than out individual stellar velocities, we consider these velocity errors to be negligable.  The error bar in the top left of the figure shows the horizontal 2006 errors as reported by `fxcor` as well as the 2005/2006 errors reported by fxcor, added in quadrature.}
\label{paperP4_check}
\end{figure*}

Due to the different wavelength coverage in the 2005 and 2006 data sets, the 2005 spectra have a smaller spectral range ($\sim$1450\AA) usable for cross correlation.  Additionally, the lower resolution of the 2005 data combined with the loss of the calcium triplet results in lower quality cross correlation fits for the 2005 data.  Errors in velocities as reported by `fxcor` for 2005 are, on average $~$12-16 km~s$^{-1}$.  The same velocity shifts applied to the 2006 data are applied to the 2005 data, for consistency, and again, a final stellar velocity was computed from the weighted average.  One field was observed in 2005 and 2006 at ({\it l,b})$=$(6$^\circ$,$-$4$^\circ$), with the same fiber configuration (i.e. the same stars in that field observed each year).  To determine the consistency of our velocity results, a star by star comparison of stellar velocities of that field was conducted (Figure \ref{paperP4_check}).  As can be seen, despite our lower resolution and less effective wavelength coverage in 2005, our stellar velocities show an offset of -2.7 km~s$^{-1}$, with an RMS scatter of 4.8 km~s$^{-1}$ about that mean.  Since the offset is less than the RMS scatter, we consider the 2005/2006 data sets to be in good agreement, and adopt ~5 km~s$^{-1}$ as our errors for individual stellar velocities.

\begin{figure*}[t]
\plotone{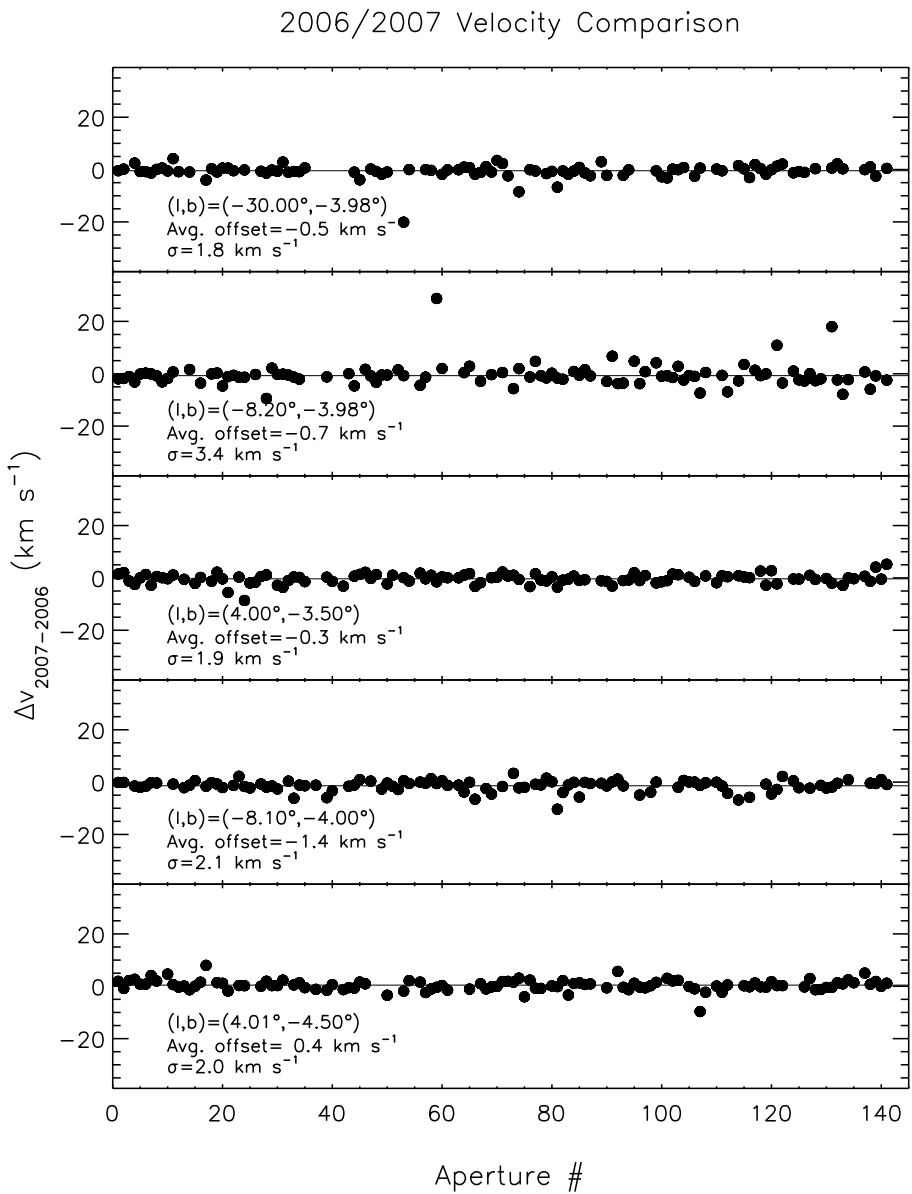}
\caption{Comparison of velocities for the 2007 observations.  For the 5 fields observed in 2007A, velocities were obtained from the four 2006 standards and the two 2007 standards.  The vertical axis are the difference in velocities and each panel shows the average offset and RMS scatter of the offset.  As can be seen, our velocities appear to be, on average, consistent to within our adopted error of 5 km~s$^{-1}$.}
\label{ladder_2007}
\end{figure*}

To check consistencies in velocity between 2007 observations and those of 2006, the 2007 fields were cross correlated with the four 2006 standards as well as the two 2007 standards, and the resulting mean stellar velocities and dispersions were compared.  Again, due to the differing wavelength coverage between 2006 and 2007, the 2006 standards correlated with the 2007 targets yielded velocities with an average error higher than the error reported by the 2007 standards.  We compare the weighted average of the two 2007 velocities with the velocities obtained from correlations of the 2006 standards and find, for the 5 fields observed in 2007, the individual stellar velocities were offset, on average, by $~$1 km~s$^{-1}$ with an avergae RMS of $~$2 km~s$^{-1}$, as can be seen in figure \ref{ladder_2007}.   Additionally, we re-observed 3 stars in 2007 that had been obtained in the 2006 dataset.  Two of these 3 stars were found to have velocities in 2007 that agree with the velocities obtained in 2006, within $~$4 km~s$^{-1}$, which is less than our adopted individual velocity error of 5 km~s$^{-1}$.  The third star was found to have a velocity difference of $~$8 km~s$^{-1}$, still in reasonable agreement with the individual errors added in quadrature.

For purposes of model comparison, we follow the lead of Beaulieu et al. (2000) and correct all heliocentric velocities obtained for the solar reflex motion using the circular motion of the LSR at the Sun as 220 km~s$^{-1}$, and the Sun's peculiar velocity relative to the LSR as 16.5 km~s$^{-1}$ toward ({\it l,b})$=$(53$^\circ$,25$^\circ$) (Kerr \& Lynden-Bell 1986, Mihalas \& Binney 1981, Beaulieu et al. 2000) by
\begin{equation}
{\it V}_{GC}={\it V}_{HC}+220sin({\it l})cos({\it b})+16.5[sin({\it b})sin(25)+cos({\it b})cos(25)cos({\it l}-53)]
\end{equation}
where {\it V}$_{HC}$ is the reported heliocentric velocity from `fxcor`.  We did not apply this transformation in Rich et al. (2007a), nor did we apply any additional color cuts (to be discussed in section 5), reporting only heliocentric velocities in that study.

\section{Color Selection and Reddening}
\begin{figure}
\centering
\includegraphics[scale=0.4]{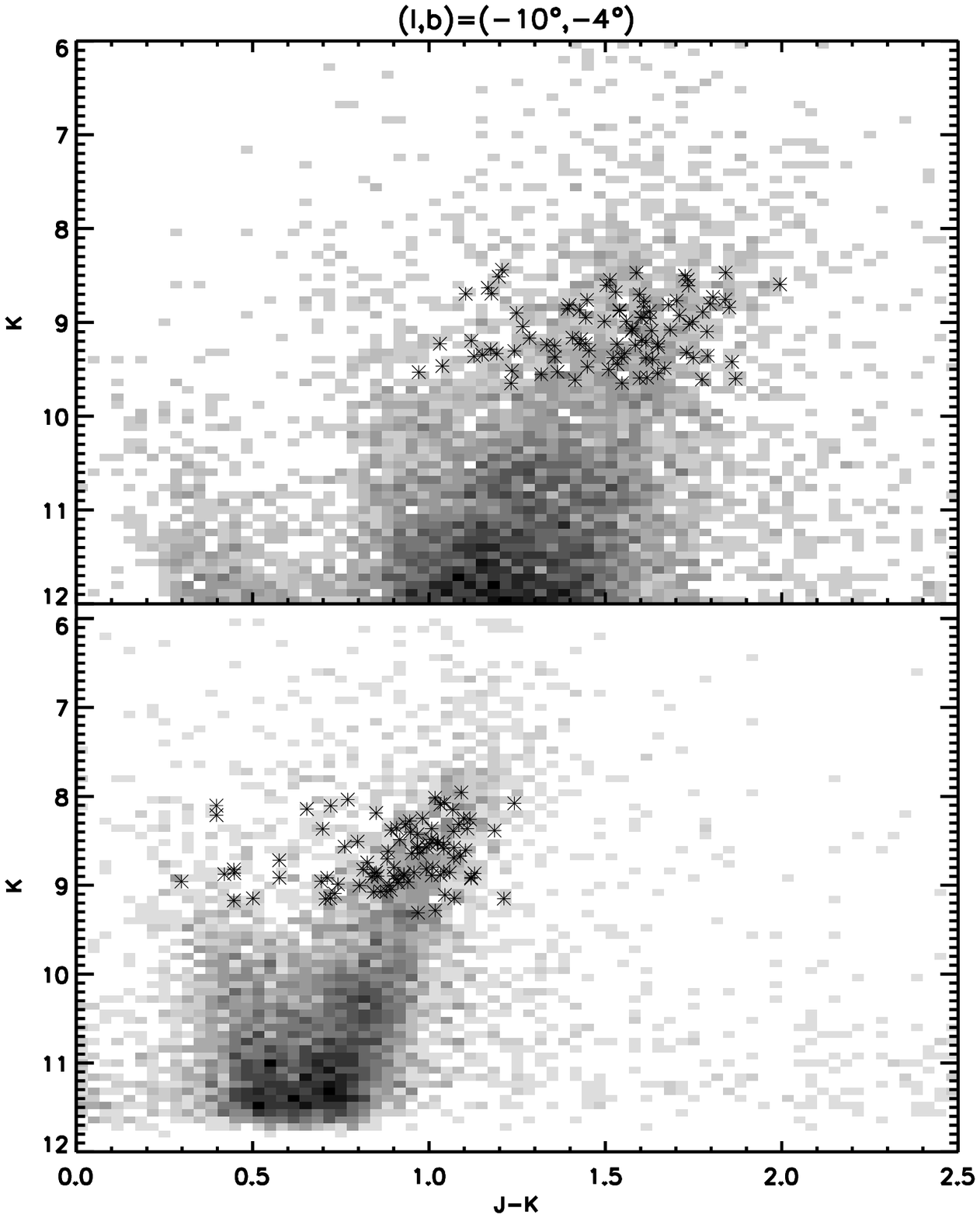}
\includegraphics[scale=0.4]{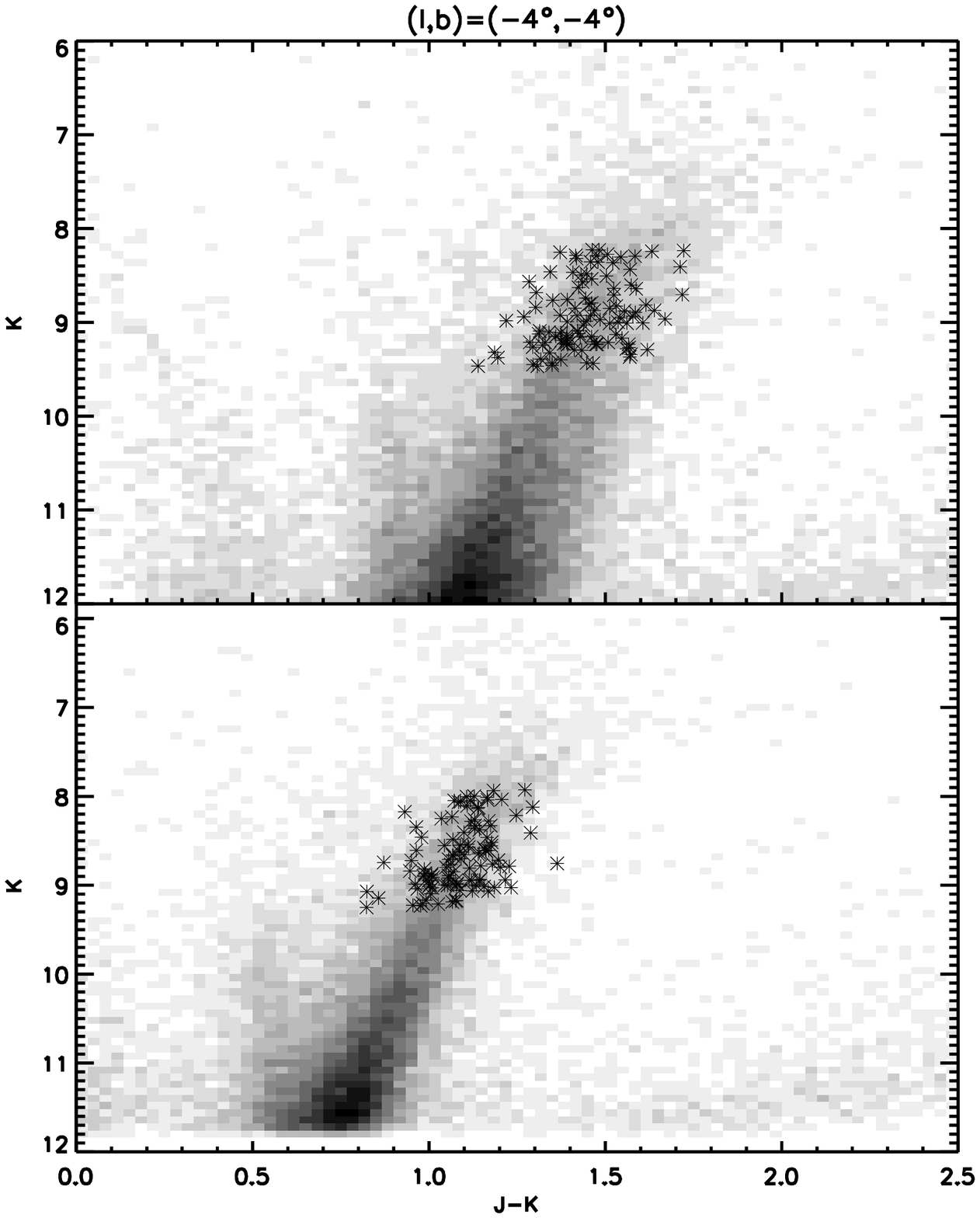} \\
\includegraphics[scale=0.4]{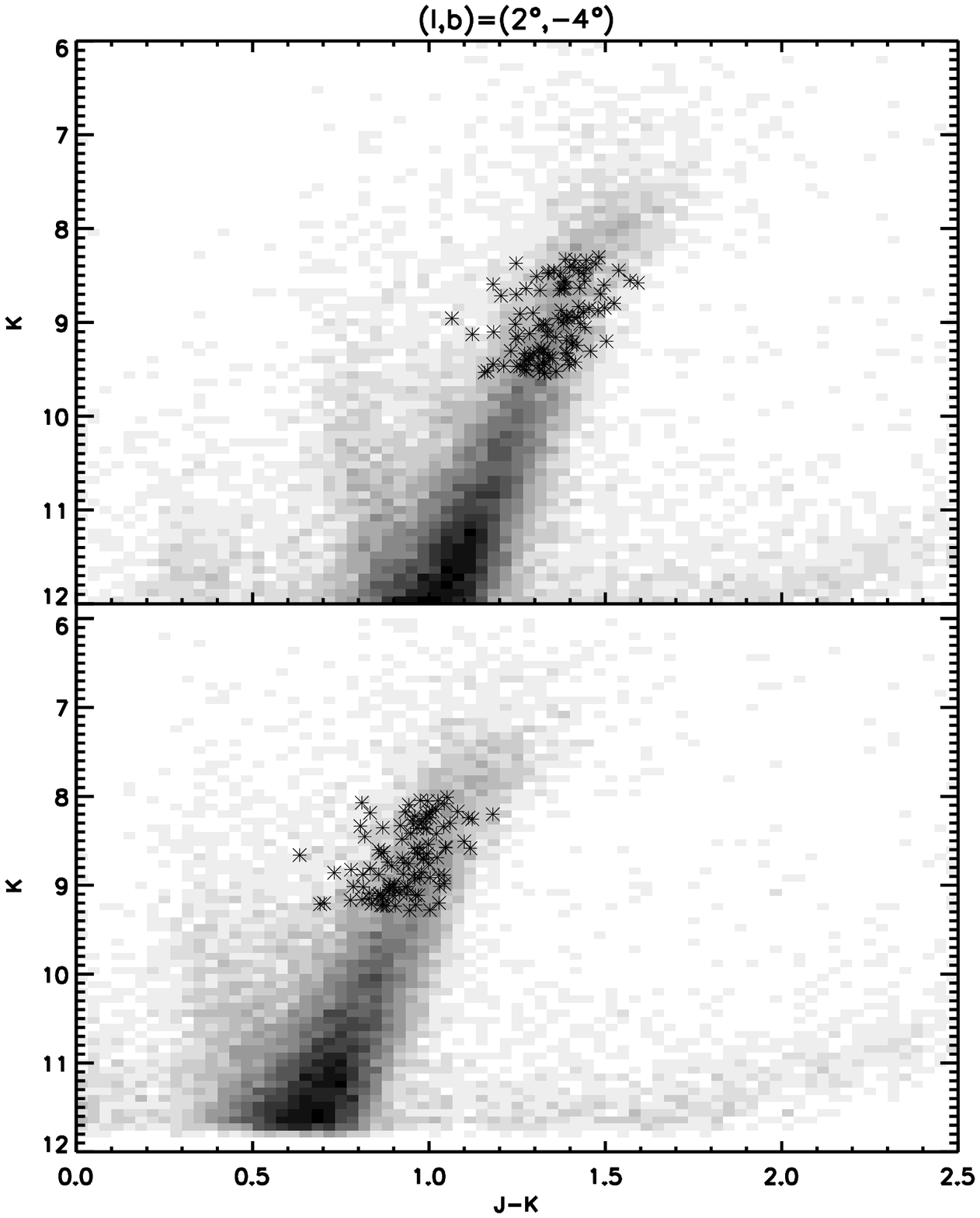}
\includegraphics[scale=0.4]{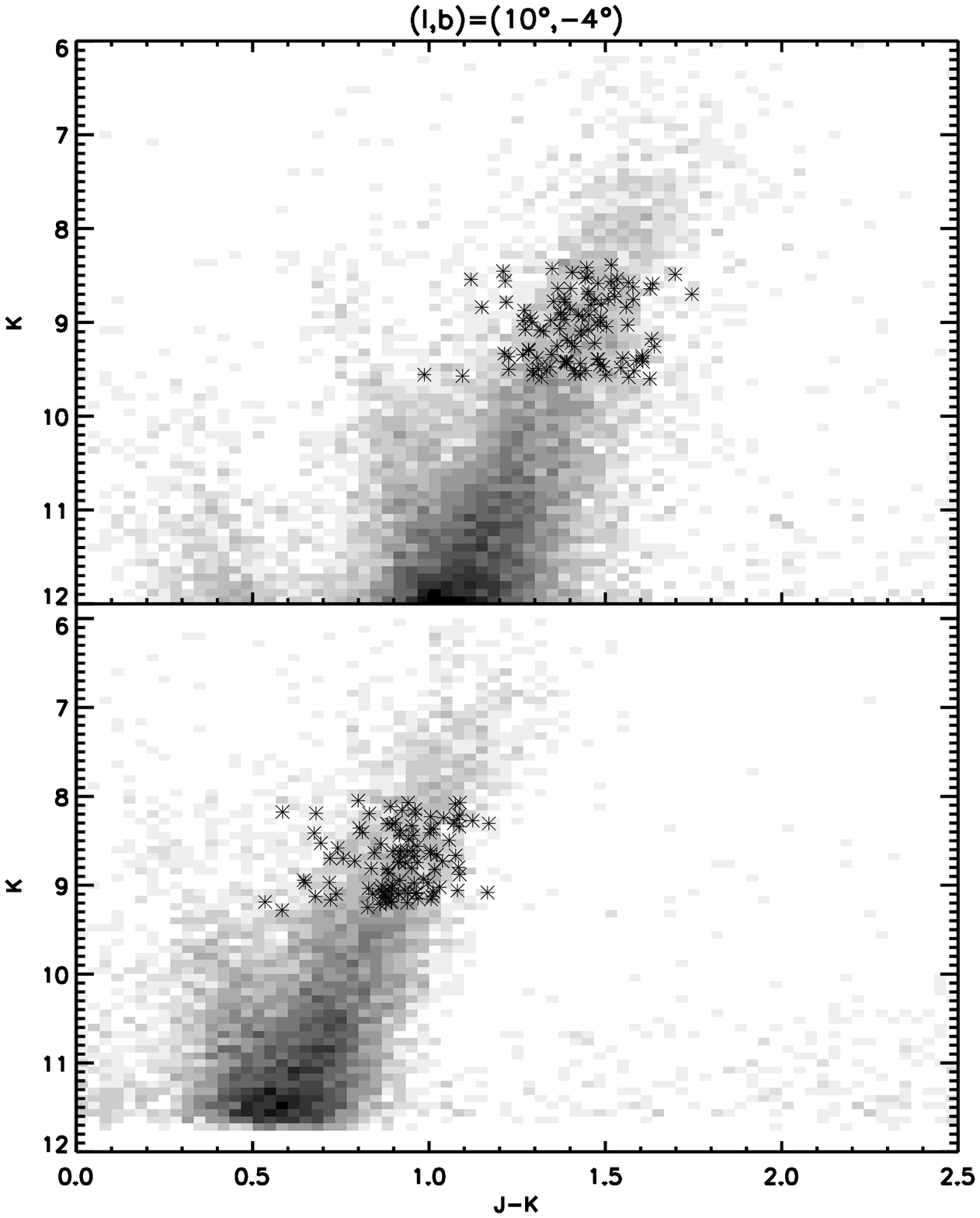}
\caption{Sample grayscale CMDs of four of the 35 BRAVA survey fields.  The upper panels of each figure show all {\it 2MASS} targets obtained in each field, with BRAVA targets marked with points.  The lower panels show the same {\it 2MASS} stars and {\it BRAVA} targets, individually de-reddened using Schelgel et al. (1998) reddening maps.  As can be seen, some fields show observed stars that appear too blue in color to be safely assured of their bulge membership.  These blue stars are omitted when calculating the field statistics (see figures \ref{total_CMD} , \ref{minor_CMD}).}
\label{de-redd}
\end{figure}

\begin{figure*}[t]
\epsscale{1.1}
\plottwo{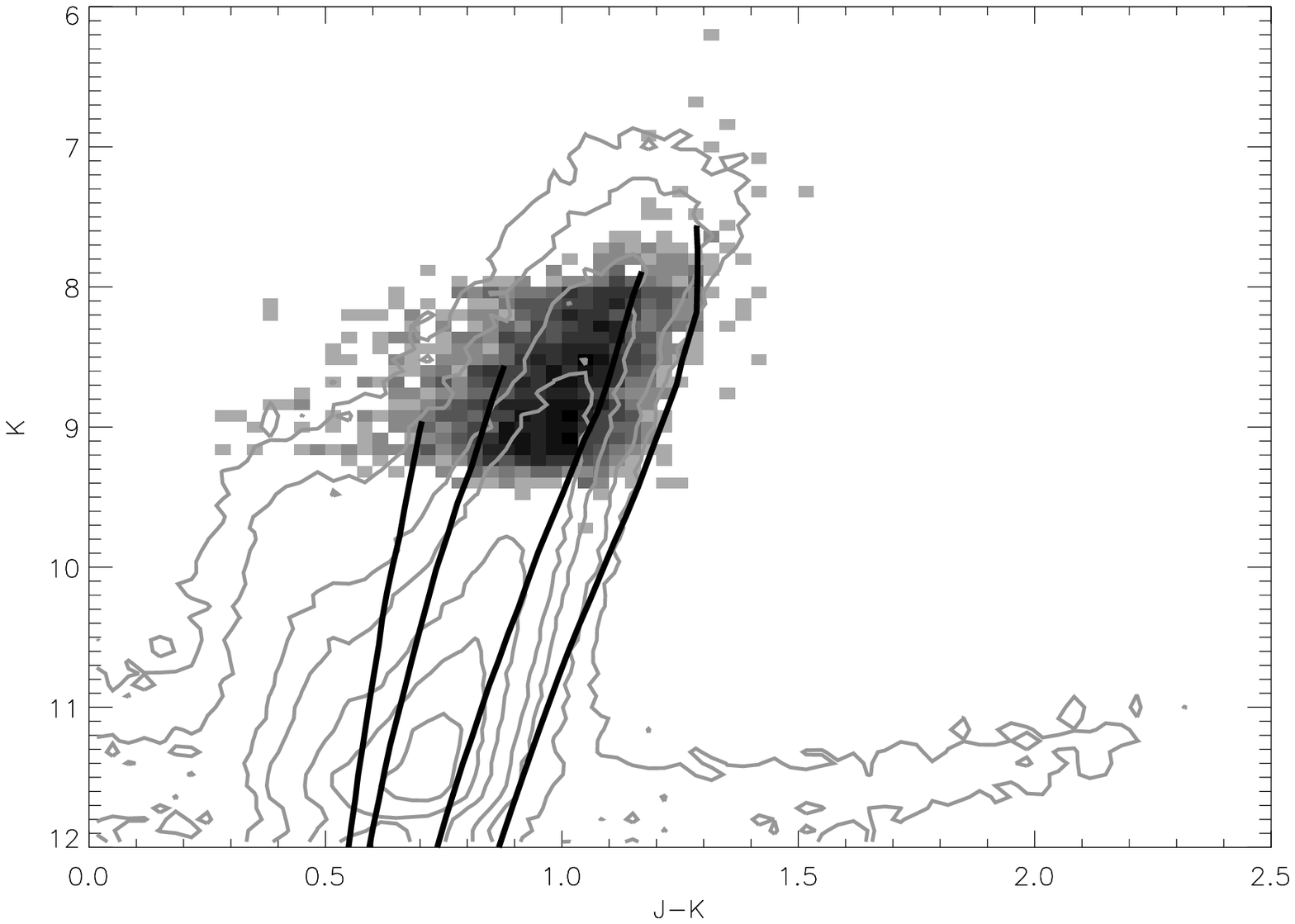}{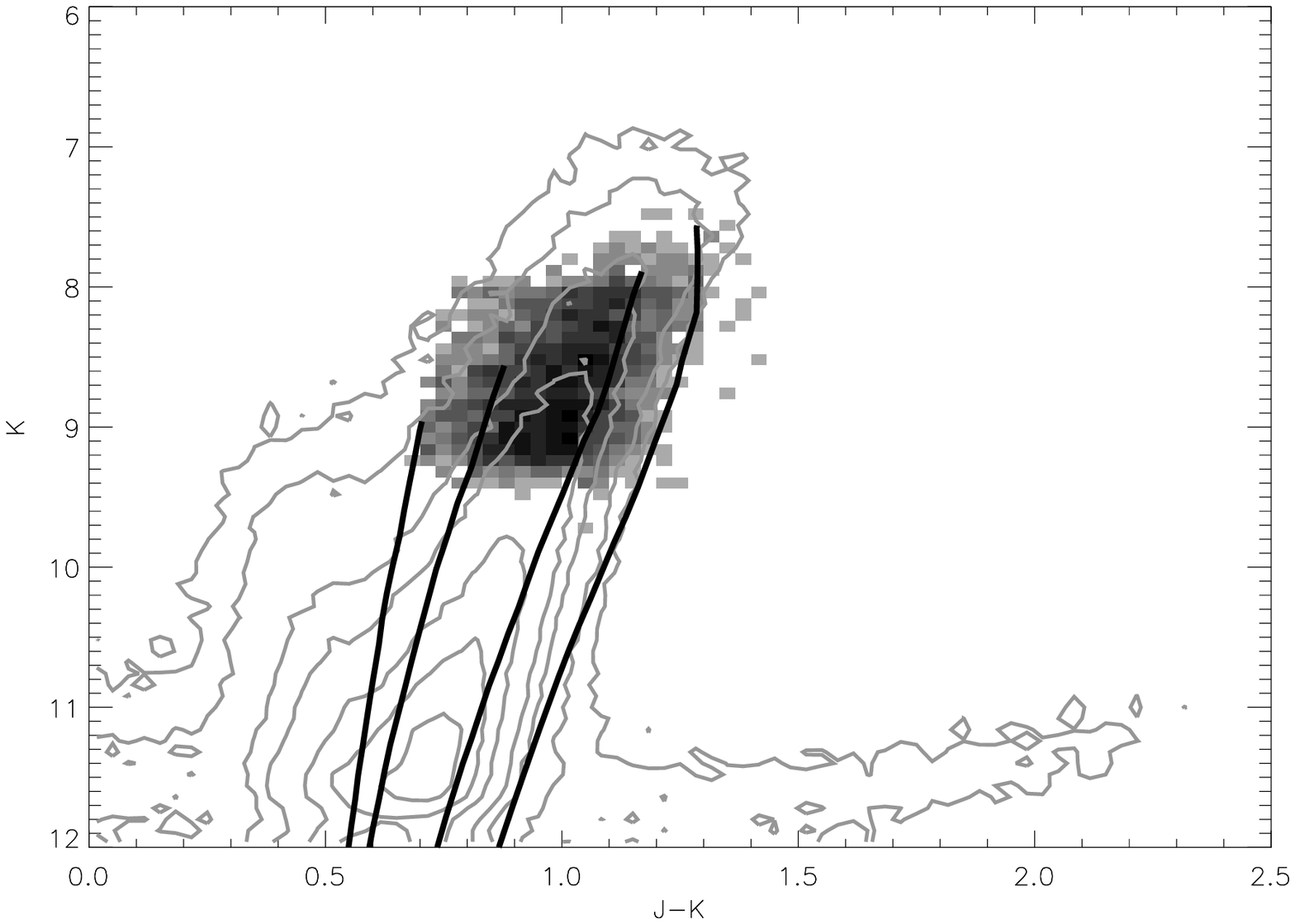}
\caption{Left: Color-Magnitude contours of the {\it 2MASS} catalog ($\sim$440,000 stars) for our fields along {\it b}$=$$-$4$^\circ$ major axis strip. Overlayed in greyscale are our observed BRAVA targets, representing 2612 stars, and isochrones for a 12 Gyr population at a distance modulus of 14.47 mag.  The isochrones, starting  on left, are for [Fe/H]$=-2.0$,$-1.3$,$-0.5$, and $+0.2$ (Marigo et al. 2008).  Right: Same as left, but showing the {\it BRAVA} targets remaining  after trimming stars bluer than the [Fe/H]$=-2.0$ isochrone and brighter than K$\sim$7.4. Out of the original 2612 targets observed along this strip, 2505 survive the cut. These 2505 targets are used in calculation of individual field statistics.}
\label{total_CMD}
\end{figure*}

\begin{figure*}[t]
\plottwo{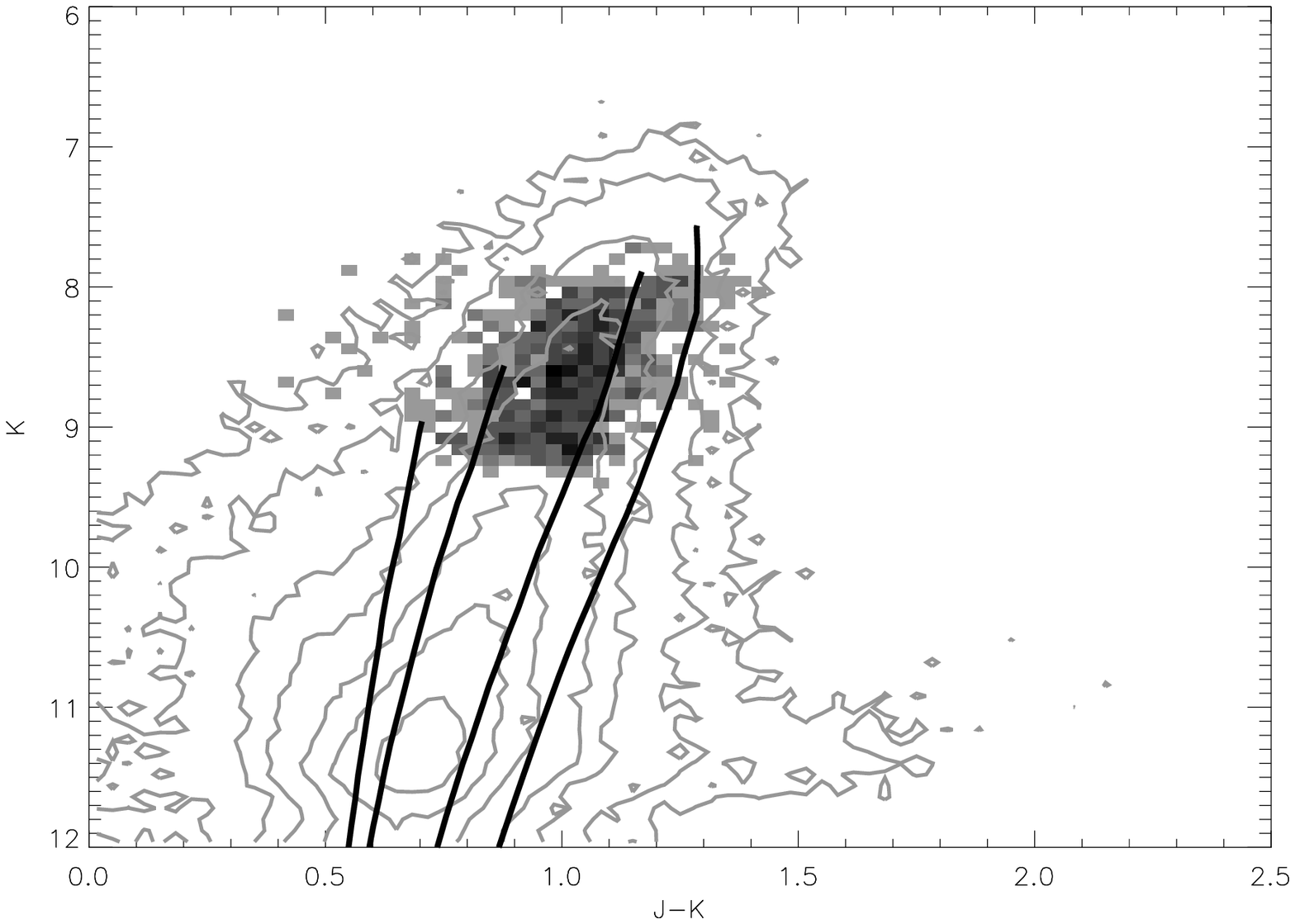}{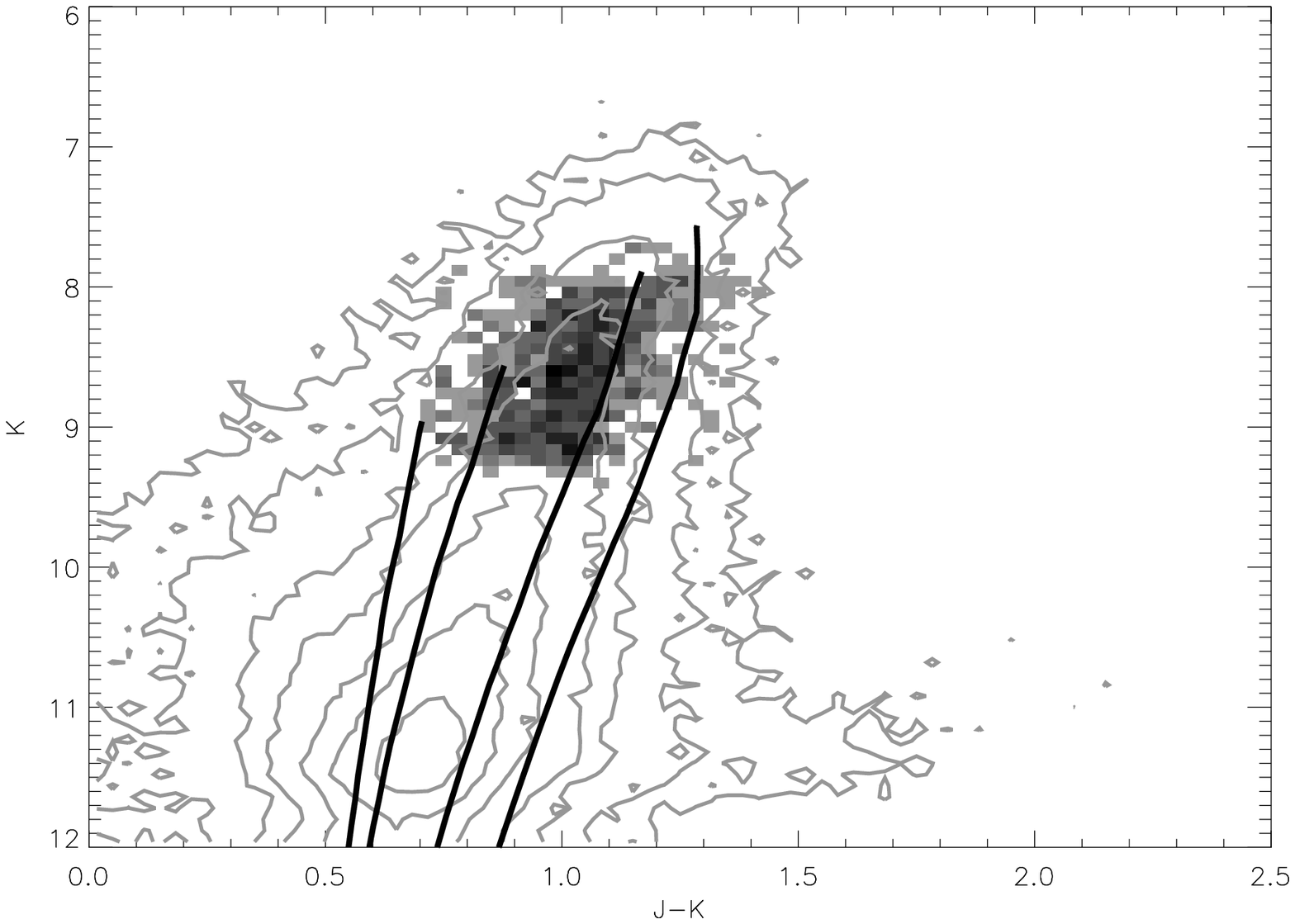}
\caption{Left: Contours of the {\it 2MASS} catalog ($\sim$290,000 stars) for our fields along $-$0.4$^\circ$$<${\it l}$<$0.0$^\circ$ minor axis strip. Overlayed in greyscale are our observed BRAVA targets, representing $\sim$850 stars, and isochrones for a 12 Gyr population at a distance modulus of 14.47 mag.  The strong effects of reddening and the uncertaintly of the Schelgel maps at {\it b}$<|5|^\circ$ can be seen in the broadening of the RGB, as compared to fig. \ref{total_CMD}  The isochrones, starting  on left, are for [Fe/H]$=-2.0$,$-1.3$,$-0.5$, and $+0.2$ (Marigo et al. 2008).  Right: Same as left, but showing the targets after trimming stars bluer than the [Fe/H]$=-2.0$ isochrone and brighter than K$~$7.4. Out of the original 850 {\it BRAVA} targets observed along this strip, 832 survive the cut. These 832 targets are used in calculation of individual field statistics.}
\label{minor_CMD}
\end{figure*}

In order to assure that our observed stars lie on the RGB, all the field CMDs were dereddened using reddening maps provided by Schlegel et al. (1998).  For each field, the {\it 2MASS} stars were dereddened as well as our observed targets (figure \ref{de-redd}).  The de-reddened fields appear reasonable and show a clearly defined ``red-edge'' of the RGB.  Figures \ref{total_CMD} and \ref{minor_CMD} show contours of the combined de-reddened {\it 2MASS} fields along the {\it b}$=$$-$4$^\circ$ major axis strip and the $-$0.4$^\circ$$<${\it l}$<$0.0$^\circ$ minor axis strip, along with a greyscale of our de-reddened observed stars and isochrones (Marigo et al. 2008).  As can be seen, the de-reddened RGB  is clearly defined, and shows the majority of our sample stars lay within a region of reasonable metallicity as expected for bulge stars (e.g. Fulbright, McWilliam, \& Rich 1996).  In order to assure that our observed targets are indeed red giants, we make a color cut omit from our sample stars bluer than the [Fe/H]=$-$2.0 isochrone (extended upwards), and stars brighter than {\it K}$\sim$7 in our calculation of the individual field statistics.  Although this reduces our observed sample by approximately 4\% (see table \ref{log1},\ref{tbl-1}), to roughly 99 stars per field, this selection can be comfortably considered to include members of the RGB of the Galactic bulge population.

For each field, mean velocities and dispersions are calculated directly from a $\sigma$-clipping algorithm, with a 6$\sigma$ cut.  None of our observed fields contained any stars with velocities deviating by more than 6$\sigma$, resulting in no stars cut from our color selected sample when calculating the statistics.  From this sigma-clipping algorithm we obtain for each field a mean velocity, velocity dispersion, and calculate the errors in those values as $\sigma$/$\sqrt{N}$ and $\sigma$/$\sqrt{2N}$, respectively.  Of course, the intrinsic dispersion of each field is given by
\begin{equation}
\sigma^{2}_{intrinsic}=\sigma^{2}_{observed}-\sum_{i=0}^{N}[error^{2}(v_{z})]/[2(N-1)]
\end{equation}
where the first term is the observed velocity dispersion of a given field and the second term represents the uncertainties in the individual stellar radial velocity (v$_{z}$) measurements, with N being the number of stars in that field.  For all fields observed, the second term is negligible and thus our observed velocity dispersions can be considered intrinsic to the field. As a check of our velocity results, the BRAVA dispersion and mean velocity measurements for the Baade's window field ({\it l,b})$\sim$(1$^\circ$,$-$4$^\circ$) are $\sigma$=112 {$\pm$} 10 km~s$^{-1}$ and V$_{mean}$=$-$5 $\pm$ 14$~$km~s$^{-1}$ and are in good agreement with that of SWC who found values of $\sigma$=113 {$\pm$} 6 km~s$^{-1}$ and V$_{mean}$=$-$4 $\pm$ 8$~$km~s$^{-1}$, which confirms the validity of our measurements.  A complete listing of our currently observed fields is shown in Table \ref{tbl-1}, including the data presented in Rich et al. (2007) updated with color cuts and galactocentric velocities. 
 
\begin{figure*}[t]
\plottwo{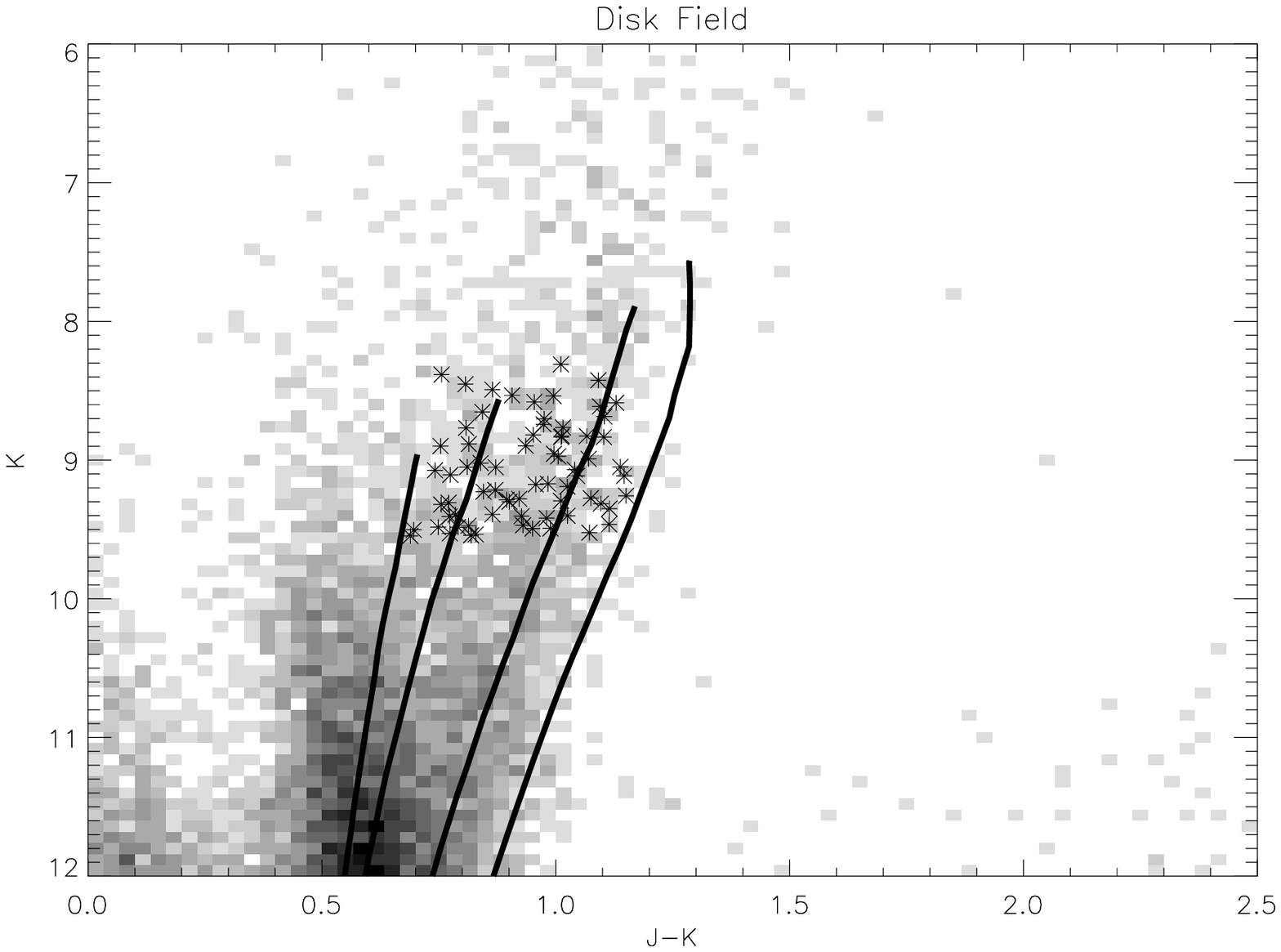}{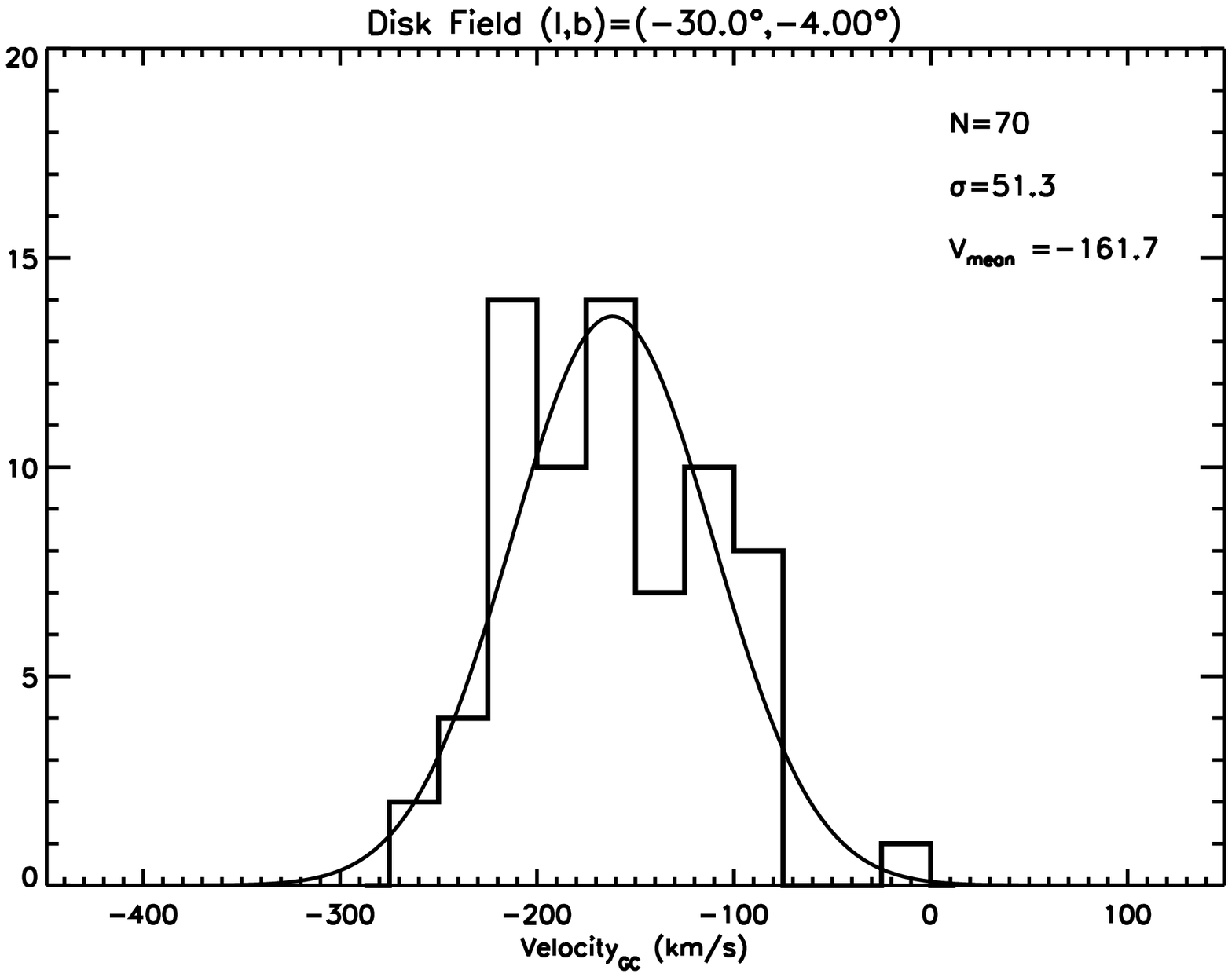}
\caption{Left: Greyscale CMD of our disk field at ({\it l,b})$=$($-$30$^\circ$,$-$4$^\circ$), with the observed targets marked as points. The same color cuts and dereddening applied to our bulge sample set are applied here.  The isochrones are the same as those described in fig. \ref{total_CMD}.  Right: Kinematic data of our disk field in the Galactocentric rest frame. Binsize is 25 km~s$^{-1}$.   Comparison with Table \ref{tbl-1} and Figure \ref{models} show this disk field to have a lower dipsersion and mean velocity that the fields in the bulge.}
\label{disk}
\end{figure*}

In order to characterize the kinematic properties of the disk, in comparison to our data, we observed a field at ({\it l,b})$=$($-$30$^\circ$,$-$4$^\circ$) in 2007.  Again, we de-reddened this field, and applied the same velocity correction for the solar reflex motion as well as the same color cuts applied to our bulge fields to obtain a mean Galactocentric velocity and dispersion (Figure \ref{disk}).  
Both the velocity dispersion ($\sigma=51\pm4$ km s$^{-1})$ and mean velocity ($-162$ km sec$^{-1}$)  contrast strongly with our bulge dispersions, which typically exceed 80 km sec$^{-1}$.   Further, we note the test applied by Sharples et al. (1990), in which those stars in Baade's Window with $I<11.8$ show $\sigma=71$ km s$^{-1}$, significantly lower than the 113 km s$^{-1}$ of the full, fainter sample.  
Our disk field, and the study of Sharples et al. (1990) do not, by themselves, prove our sample free of disk contamination.   
However, the disk field departs so strongly from our rotation and dispersion profiles (Figure 12) that we suspect disk contamination is not an important issue.  Composition measurements and modeling of the CMD, as well as comparison of our data to the Besancon galaxy model, will constrain further the level of disk contamination.

\section{Results and Analysis}

\begin{figure*}[t]
\epsscale{1.2}
\plottwo{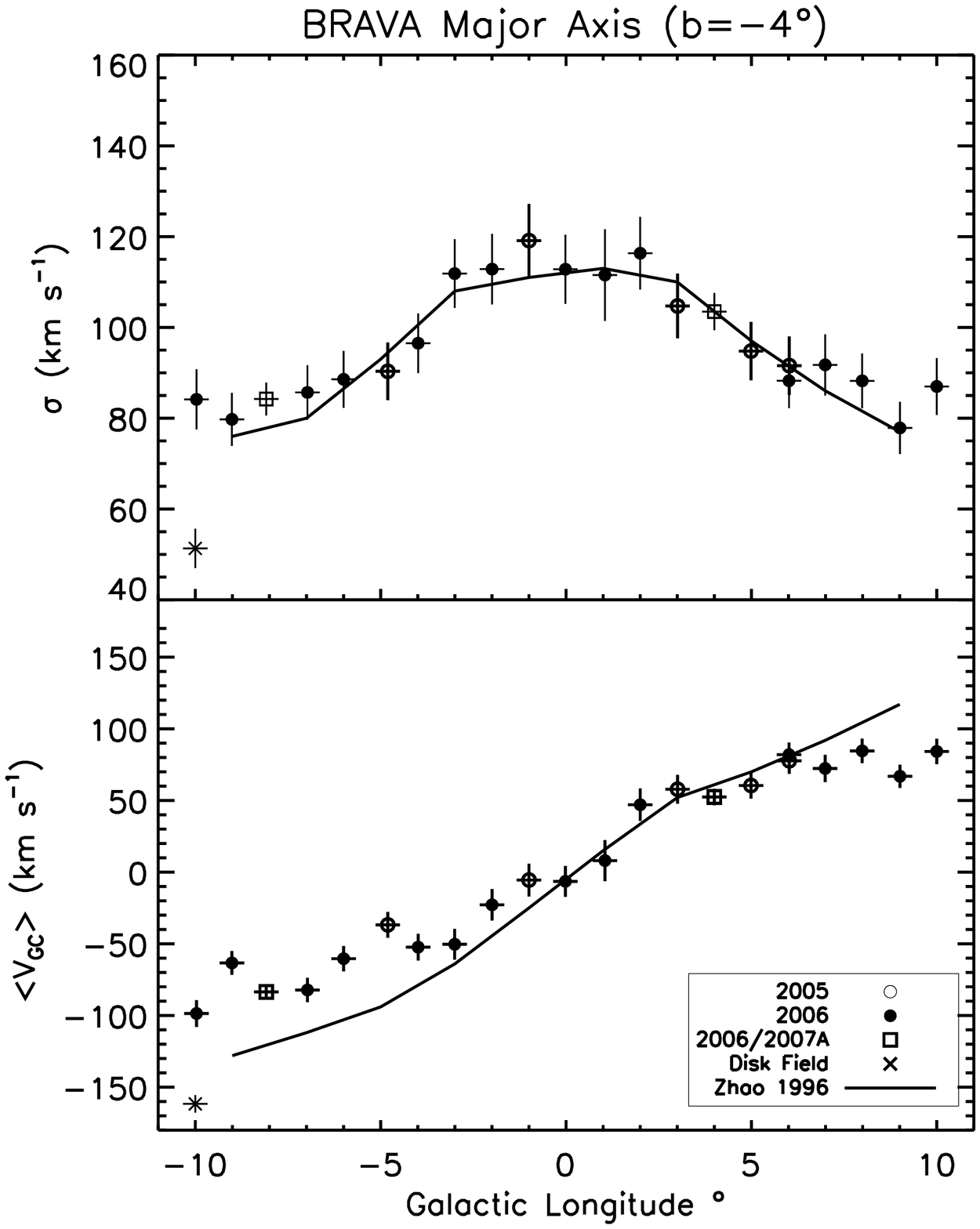}{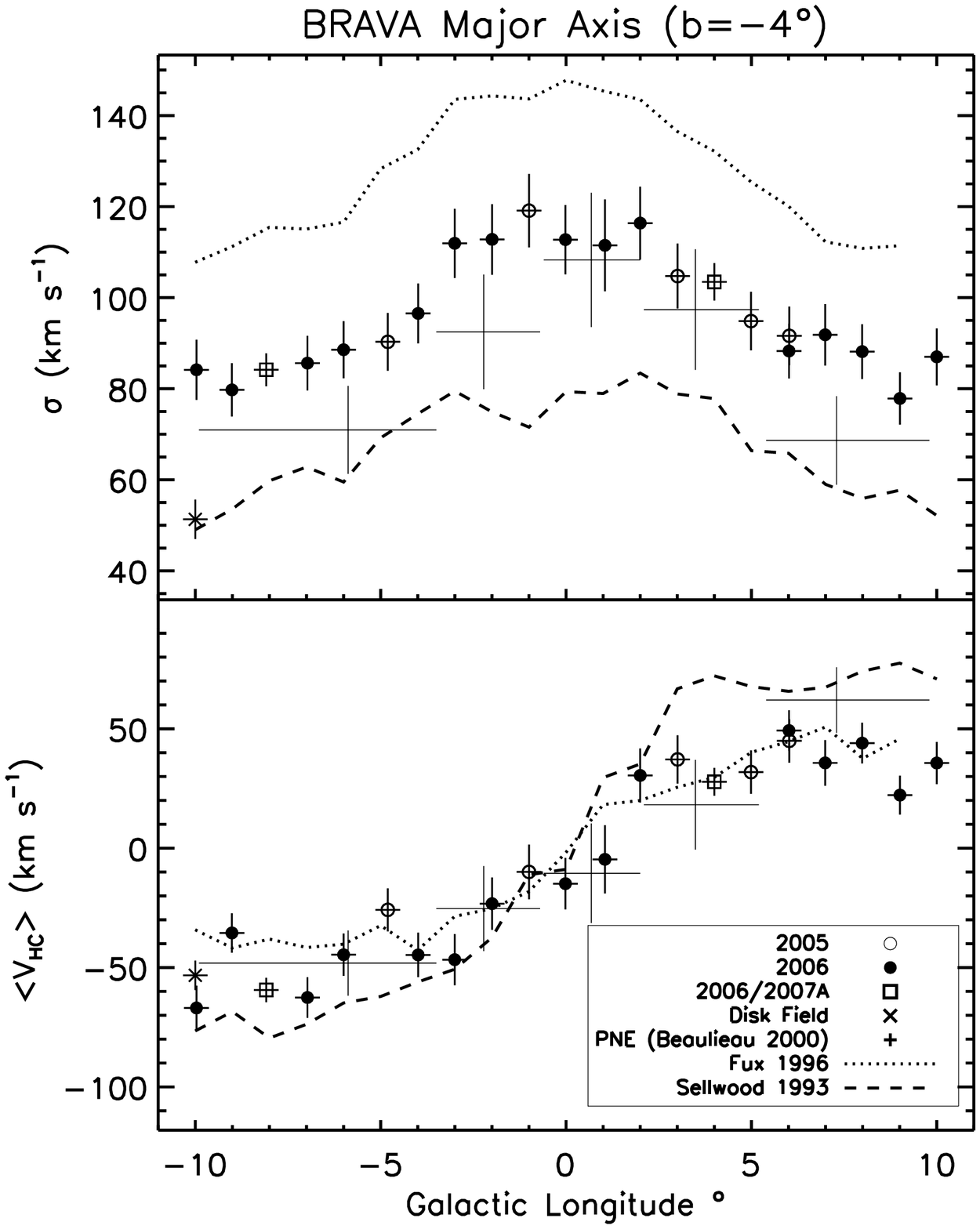}
\caption{Left: Velocity dispersion profile and rotation curve of bulge major axis, {\it b}$=$$-4$$^\circ$, in galactocentric velocity; The solid line indicates the model of Zhao (1996).  Zhao's model satisfactorilly predicts the dispersion, but shows a more rapid rotation that what is observed. Our disk field at ({\it l,b})$=$($-$30$^\circ$,$-$4$^\circ$) has been plotted to show the lower dispersion from a disk field, as expected.  Right: Major axis velocity dispersion profile and rotation curve, in Heliocentric velocity. Also indicated are the PNe data of Beaulieau et al. (2000) ($"+"$ symbols) and the models of Fux (1997) (dotted line), and Sellwood (1993) (dashed line).  Our velocity dispersion are higher than those of the PNe sample, while our rotation is consistent with the PNe.  As a homogenous population, the PNe may suffer some disk or forgeground contamination.  For the first time, our densely sampled fields follow the qualitative inflections of the N-body dispersion and rotation profiles of Fux (1996) and Sellwood (1993) quite well.}
\label{models}
\end{figure*}

\begin{figure*}[t]
\epsscale{0.8}
\plotone{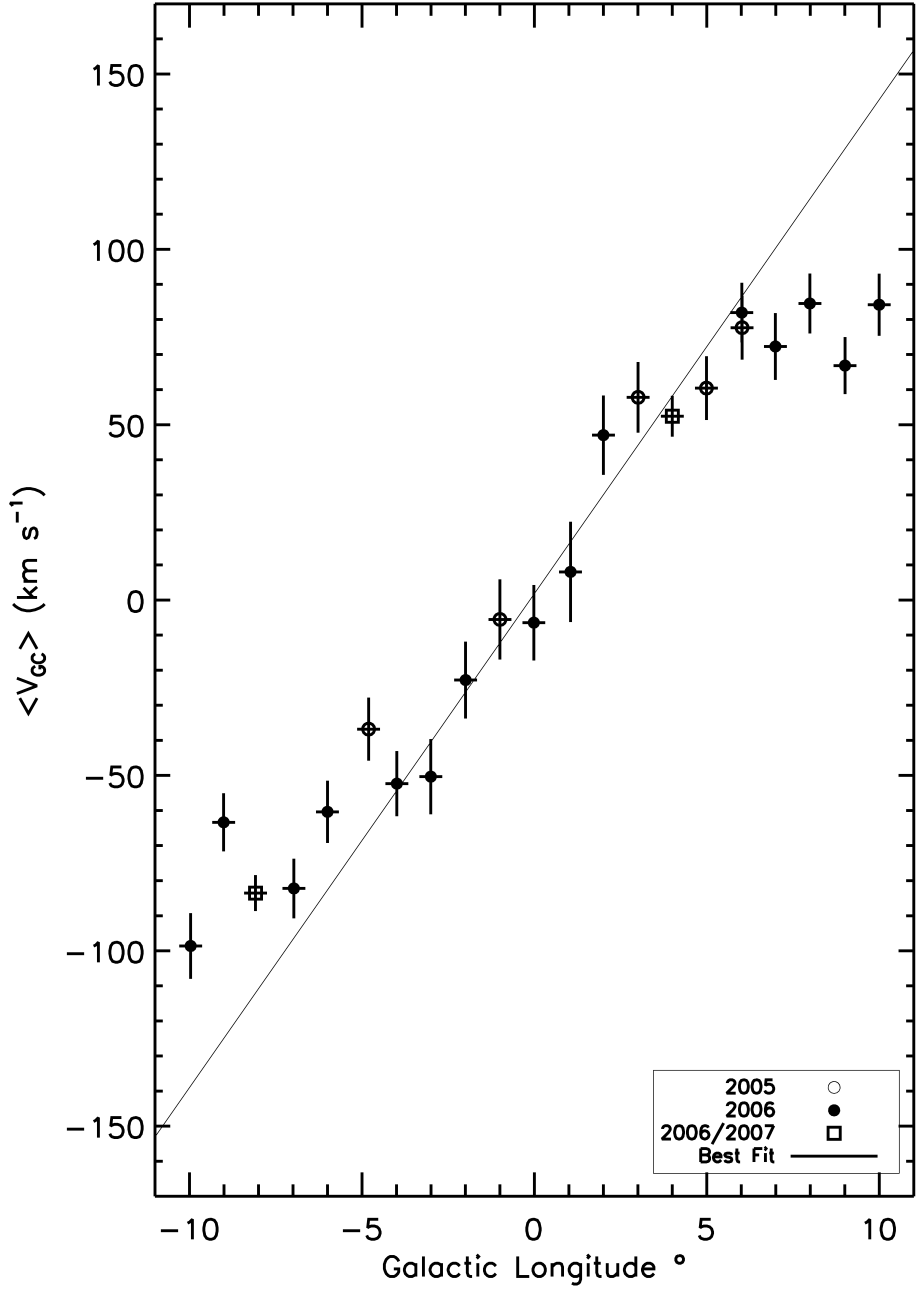}
\epsscale{0.5}
\centering
\caption{Best fit line of the central 8$^\circ$ of the Galactic bulge.  The line shows an upper limit of the slope of V$_{Rot.}$$=$100 km~s$^{-1}$~kpc$^{-1}$.  The break from solid body rotation is evident at $|${\it l}$|$$\sim$4$^\circ$.}
\label{slope}
\end{figure*}

Figure \ref{models} shows the {\it BRAVA} velocities and dispersions compared to the predictions of the Zhao (1996) model, in Galactocentric velocity rest frame. The Zhao model is a 3D steady-state model which uses a generalized Schwarzchild technique, consisting of orbital building blocks within a rapidly rotating bar potential, with corotation at 3.3 kpc, and a Miyamoto-Nagai (MN) disk potential (see Zhao 1996 for details).  In this model, roughly half of the mass of the bar consists of stars on direct regular orbits which form the rotational support of the bar, and the remaining half in irregular and retrograde orbits. The data agree with the Zhao model dispersion, but shows a significant deviation from the predicted rotation curve at large Galactic longitude.  Solid body rotation claimed by numerous previous studies (Menzies 1990; Izumiura et al. 1995) is not apparent in the {\it b}$=$$-$4$^\circ$ major axis strip; after reaching an amplitude of $~$75 km~s$^{-1}$, the rotation curve flattens beyond $|${\it l}$|$$\sim$4$^\circ$, coresponding to a projected distance of $\approx$550 pc from the central minor axis for R$_{o}$$=$8kpc.  Furthermore we find that the upper limit for rotation of the bulge, in Galactocentric rest frame velocity, is V$_{Rot.}$$\lesssim$100 km~s$^{-1}$~kpc$^{-1}$ for $|$ {\it l} $|$$\leqslant$4$^\circ$, and flattens beyond that at a value of roughly 75 km~s$^{-1}$ (Figure \ref{slope}).  Our upper limit for rotation value is significantly larger than the previously reported value of V$_{Rot.}$$=$70 km~s$^{-1}$~kpc$^{-1}$ obtained from PNe (Beaulieu et al. 2000). Interestingly, this velocity flattening is also observed at the same projected distance in the proper motion studies of Clarkson et al. (2008, private communication), perhaps indicating a change in the density profile.  Also plotted in Fig. \ref{models} is the mean velocity and dispersion for our observed disk field at ({\it l,b})$=$($-$30$^\circ$,$-$4$^\circ$).  As can be seen, our disk field shows lower dispersion and significantly greater rotation speed, than any of our bulge fields; this gives us added reassurance that our sample is not contaminated by foreground/disk stars.  

It is also useful to compare BRAVA data to the PNe from the survey of Beaulieu et al. (2000), and to follow their example by comparing {\it BRAVA} data to other N-body models.  Two such models are plotted in Figure \ref{models}, in heliocentric velocity.  The models of Fux (1996) and Sellwood (1993) are both N-body bars formed from initially unstable disks, investigating the initial formation of the bulge (see Beaulieu et al. 2000 for details).  Figure \ref{models} shows that none of the models are satisfactory in predicting both dispersion {\it and} rotation, suggesting that the data are challenging to fit with either disk-instability formed bars or Schwarzchild models with fixed potentials (Zhao 1996).  However, there is striking agreement in the qualitative inflections of the Fux (1996) dispersion profile and our bulge sample, differing only in a relative shift in the scale of the dispersion, while the model rotation curve is in excellent agreement with the data.  Similarly, the Sellwood (1993) model shows qualitative consistency with the observed data; although differing in amplitude, the model shows the same flattening seen in the observed rotation curve.  Due to the limited number of PNe at {\it b}$=$$-$4{$^\circ$} and {\it l}$=$$|$10$|${$^\circ$},  we accept those in the range of $-8${$^\circ$}$<$ {\it b} $<$$-3${$^\circ$}.  Within that longitude/latitude range there remain only 133 PNe for comparison with our data; the PNe are then binned so that $\sim$27 PNe reside in each bin (Fig. \ref{models}).  Considering the less secure distances and the assignment of population of PNe, the agreement is good and is not consistent with solid body rotation for the bulge.  

\begin{figure*}[t]
\epsscale{1.0}
\plotone{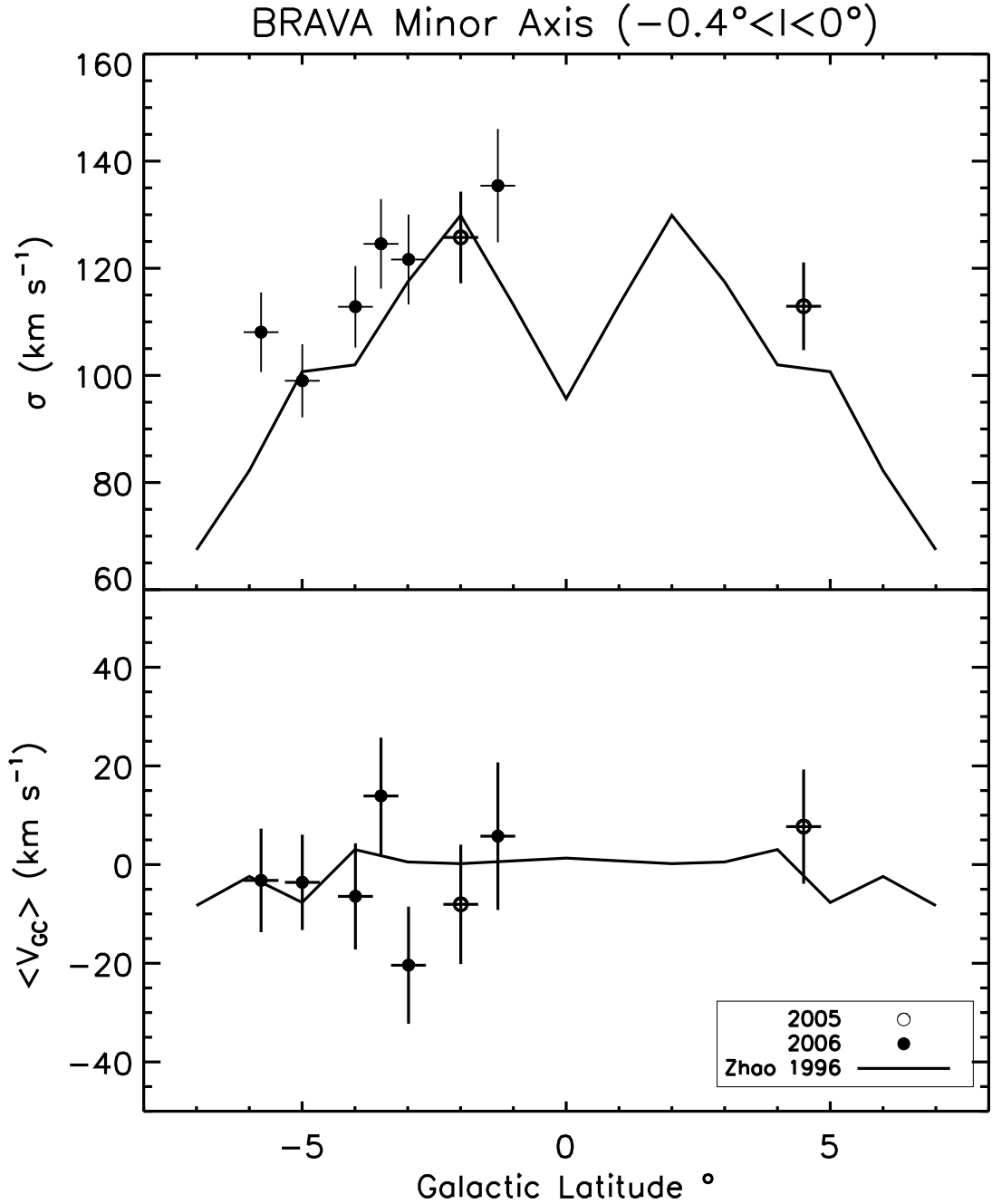}
\caption{Comparison of data with the Zhao (1996) model of the Galactic bulge minor axis ({\it b}$=$0$^\circ$,$-$0.4$^\circ$$<$ {\it l} $<$0$^\circ$).  
Top- The mean velocity is consistent with the model in that it shows no appreciable minor axis rotation. Bottom-The velocity dispersion is consistent with the trend of the model at negative latitudes, although more data is needed at positive latitudes.}
\label{minor}
\end{figure*}

{\it BRAVA} also obtained a minor axis strip of the bulge and has found no evidence of minor axis rotation (Figure \ref{minor}), although more data are needed at positive latitudes (we note that the much greater extinction at positive latitudes will make this effort challenging).  It is also noteworthy that our rotation curve now agrees with the two points observed by Minniti et al. 1992 (see discussion in Beaulieau et al. 2000), but we emphasize that for ({\it l,b})=(12$^\circ$, 3$^\circ$) Minniti et al. (1995) describe the field as "heavily contaminated by disc stars" and we consider any agreement between this field and {\sl BRAVA} to be fortuitous; the connection between the bulge and fields with $|l|>10^\circ$ demands more investigation (see Figure 7 of Beaulieau et al. 2000).  Indeed, inspection of figure 7 of Beaulieu shows a break in velocity and dispersion beyond $|l|>10^\circ$, suggesting two different populations being sampled.  Reassuringly, our disk field at ({\it l,b})$=$($-$30$^\circ$,$-$4$^\circ$) matches the PNe data quite well and illustrates the importance of gathering more data to investigate the kinematics of fields between 10$^\circ$$<|$l$|<$30$^\circ$.  In our opinion, our {\it 2MASS} selection method is better able than optical colors to defeat the reddening and to separate cleanly the bulge giants from foreground red giant and clump stars.  At ({\it l,b})$=$(12$^\circ$,3$^\circ$), it would be critical to have sampled adjacent fields and to have established a connection between this field and the higher surface brightness, inner bulge fields.  At this low latitude, disk contamination becomes a concern in any optically selected sample.  

\begin{figure}
\centering
\includegraphics[scale=0.4]{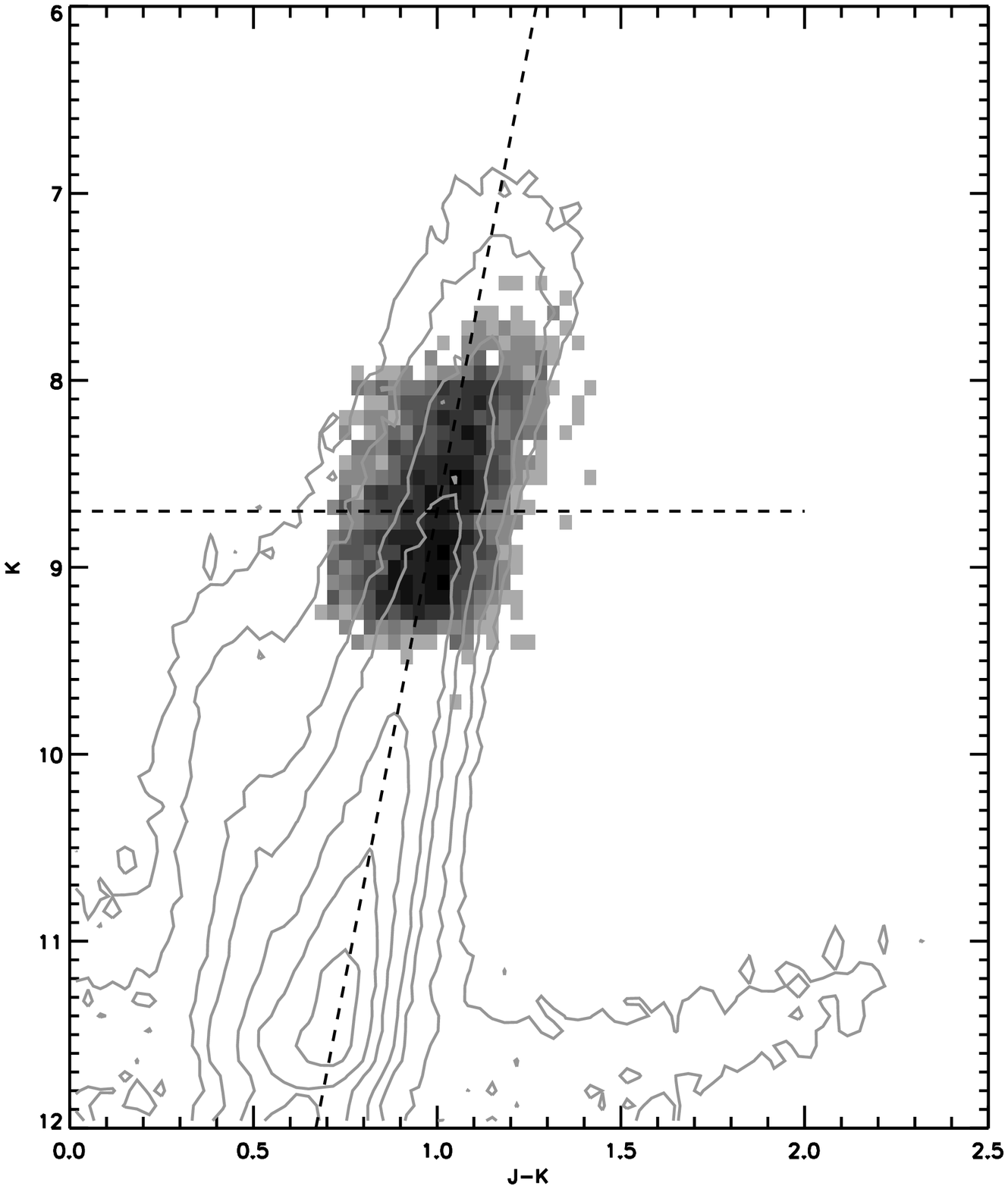}\\
\includegraphics[scale=0.4]{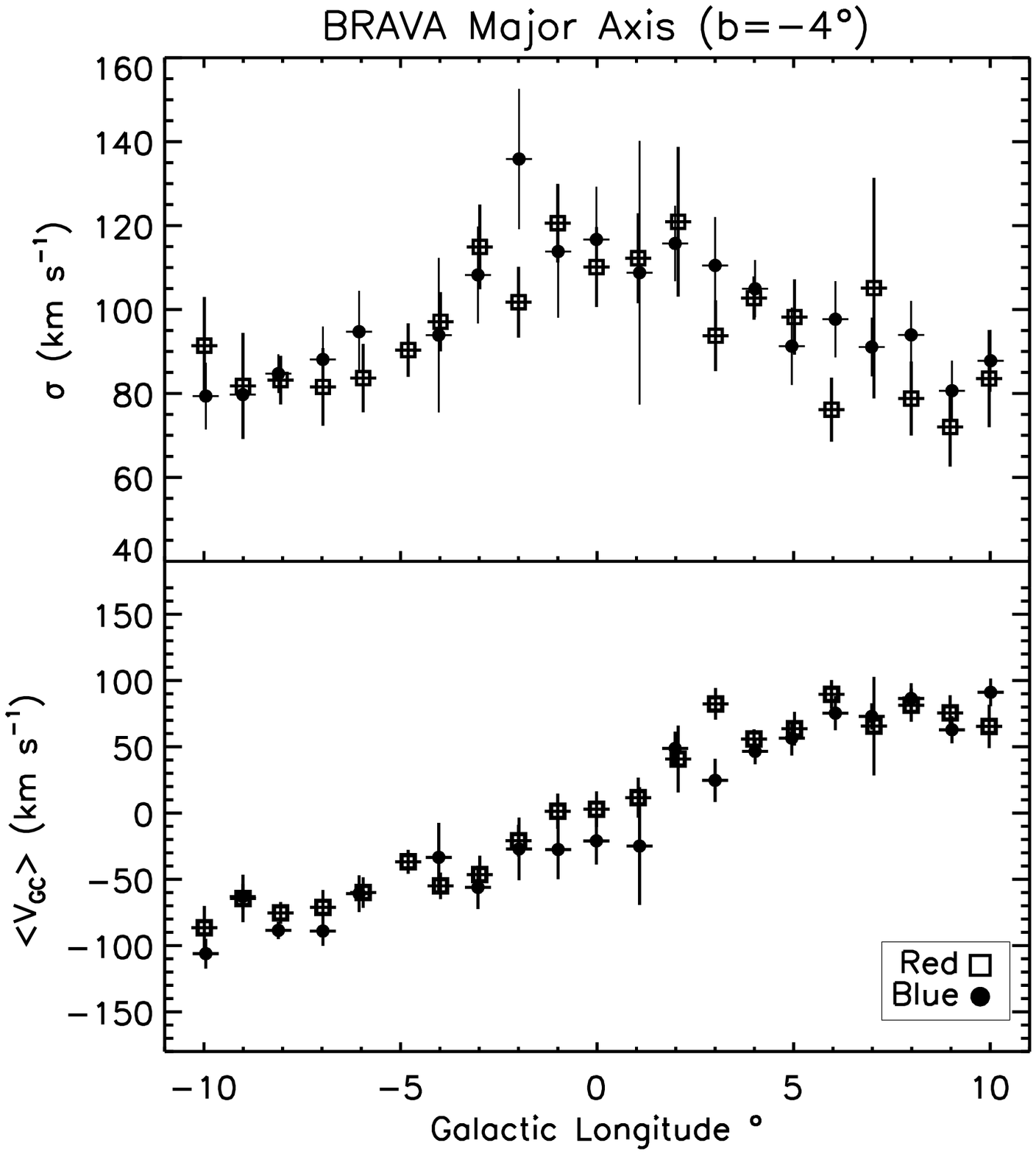} 
\includegraphics[scale=0.4]{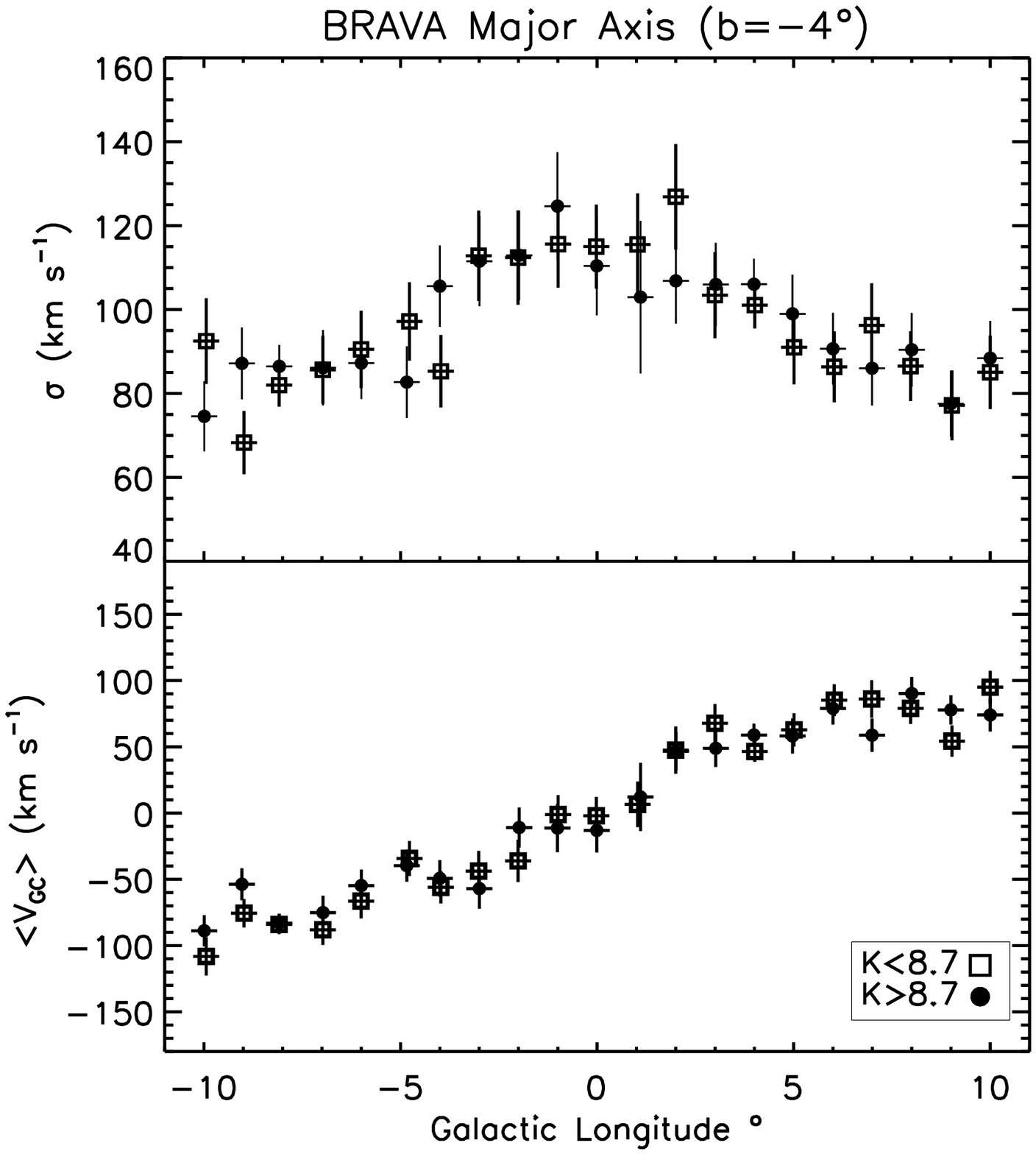}
\caption{Kinematics of the bulge major axis, separated by color/magnitude.  The top plot shows the total de-reddened {\it 2MASS} CMD of the bulge in countour, with our observed targets in greyscale.  The dashed lines represent our cuts for metallicity and brightness.  As can be seen, there appears to be no bias in our kinematics based on brightness.  The color cut offers a {\it hint} that the red (metal rich) population has a slightly lower velocity dispersion in some fields; followup with larger samples and actual metallicity measurement is required to confirm or refute this.}
\label{divide}
\end{figure}

To check for metallicity and/or magnitude bias in our survey, we have divided our de-reddened sample by magnitude (K$<$8.7 and K$>$8.7) as well as color (see Figure \ref{divide}).  Dividing our sample by magnitude seems to have little effect upon the kinematics of the major axis strip, further reinforcing our belief that our sample does not experience significant contamination from other foreground stellar populations, to first order.  Dividing our sample by color shows little effect as well, however, there appears to be a  general trend indicating that the more metal rich giants show a lower dispersion in each individual field.  Of course, by dividing our sample, we reduce the statistical significance of this apparent trend.  We plan to measure the spectroscopic metallicities of our sample.

\begin{figure*}[t]
\epsscale{1.0}
\plotone{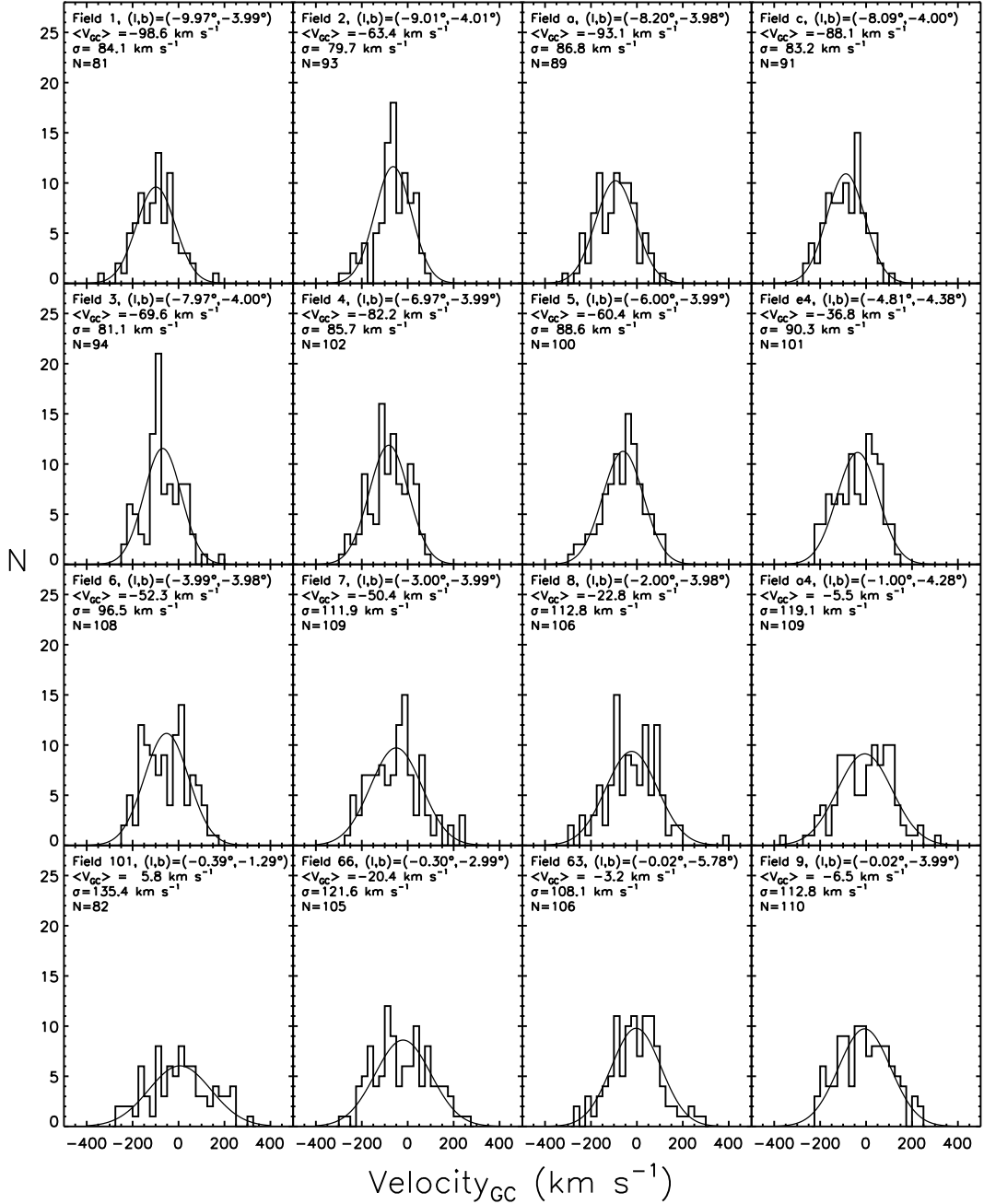}
\caption{Presentation of all bulge field Galactocentric velocity distributions.  Overlaid on each plot is a Gaussian with parameters derived from the IDL $\sigma$-clipping algorithm 'meanclip' used to calculate field statistics.  As can be seen, several fields display a departure from a normal distribution, most notably fields 2,3,7,8, and 12.  However, we stress that these departures can be a result of choice of bin size.  Fields 3 and 12 were chosen for follow-up observation in 2007; the deviation in those fields were not statistically significant (see fig. \ref{streams} for details).  Not shown are the velocity distributions of our disk field (fig. \ref{disk}), or field P4-2005 which has the same stars observed as field P4-2006.  Bin size is 25 km~s$^{-1}$.}
\label{all_hist}
\end{figure*}

\begin{figure*}[t]
\epsscale{1.0}
\plotone{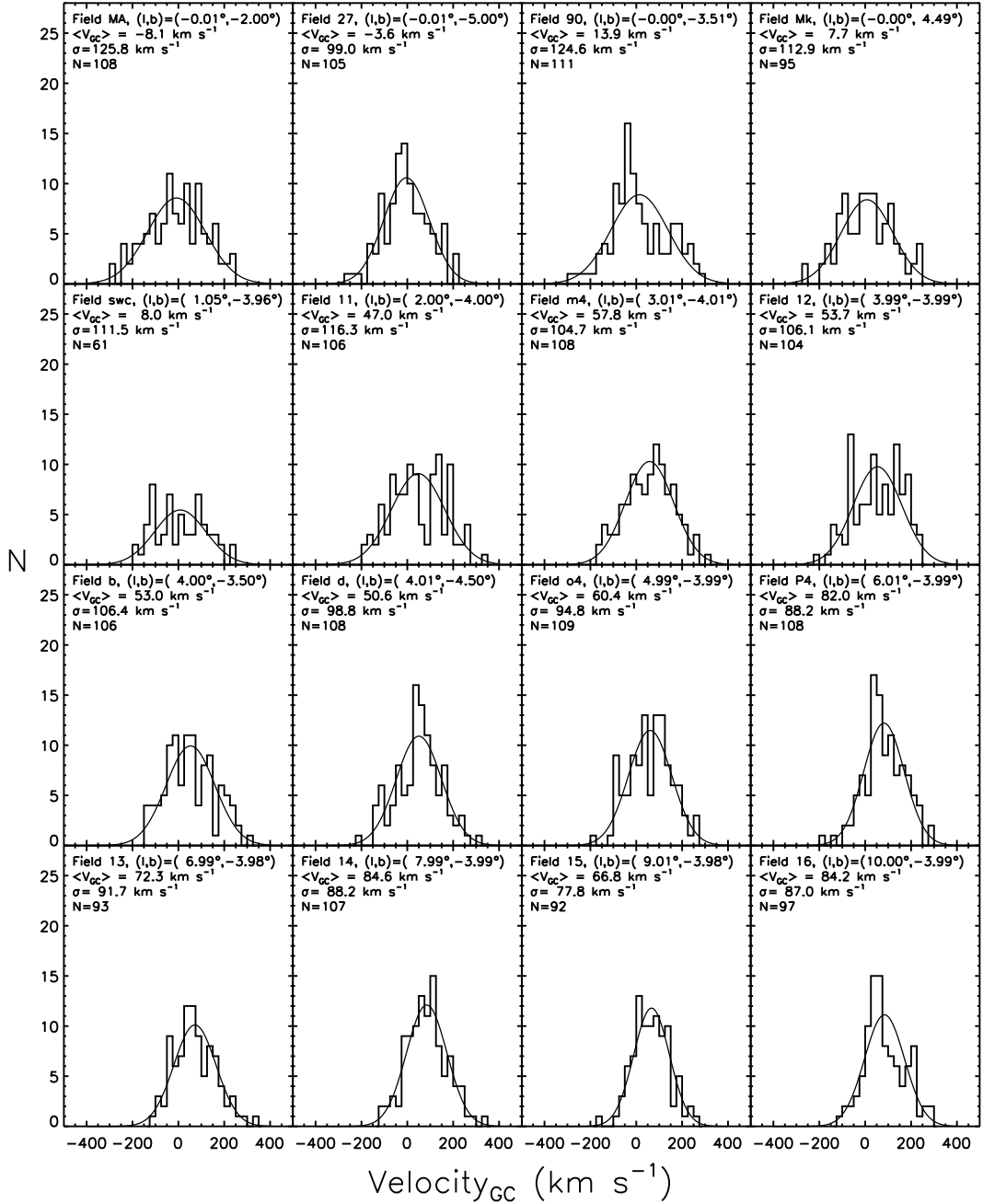}
\caption{Bulge line of sight velocity distributions cont'd. (see fig. \ref{all_hist})}
\label{all_hist_1}
\end{figure*}

\begin{figure*}[t]
\epsscale{1.0}
\plottwo{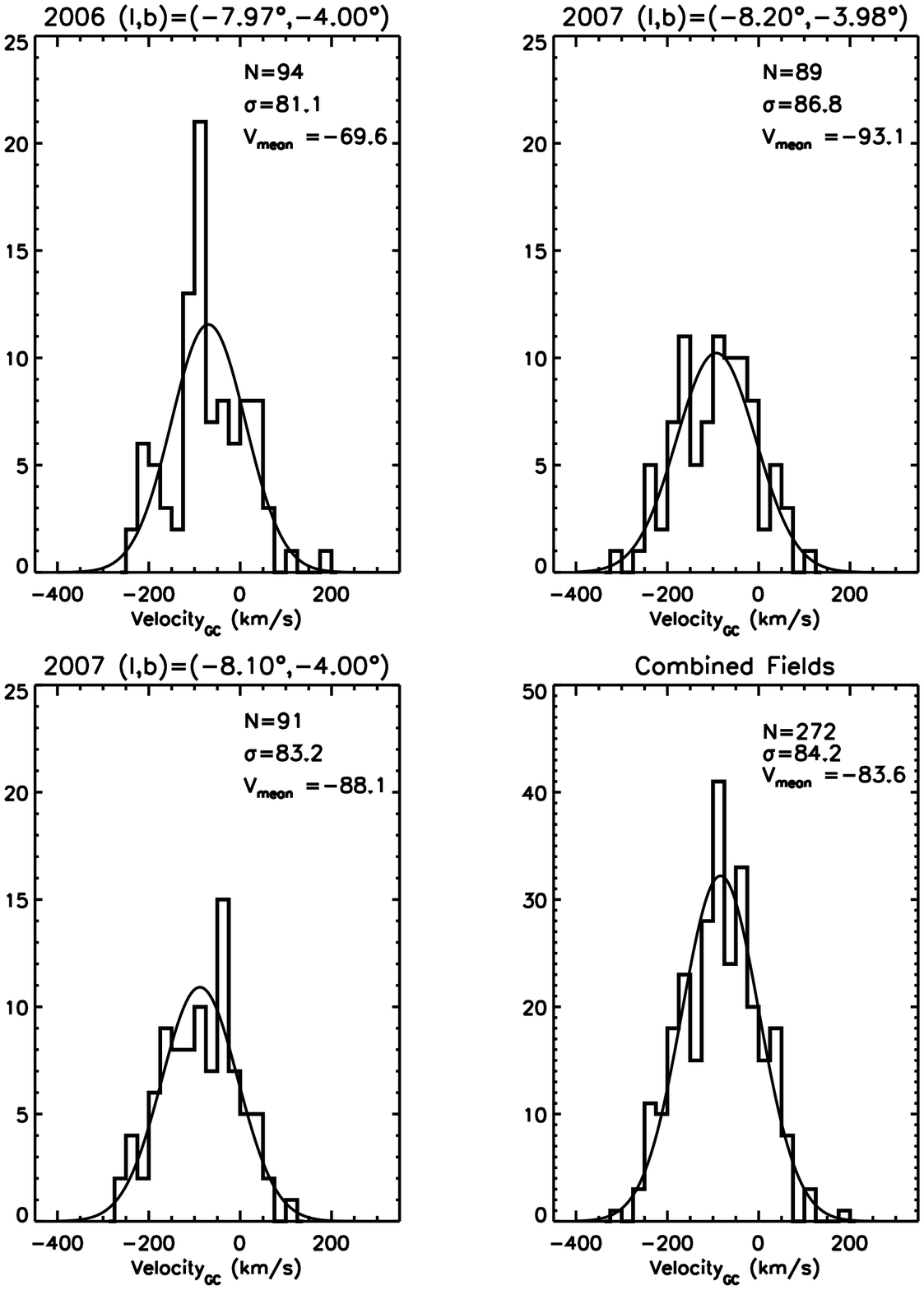}{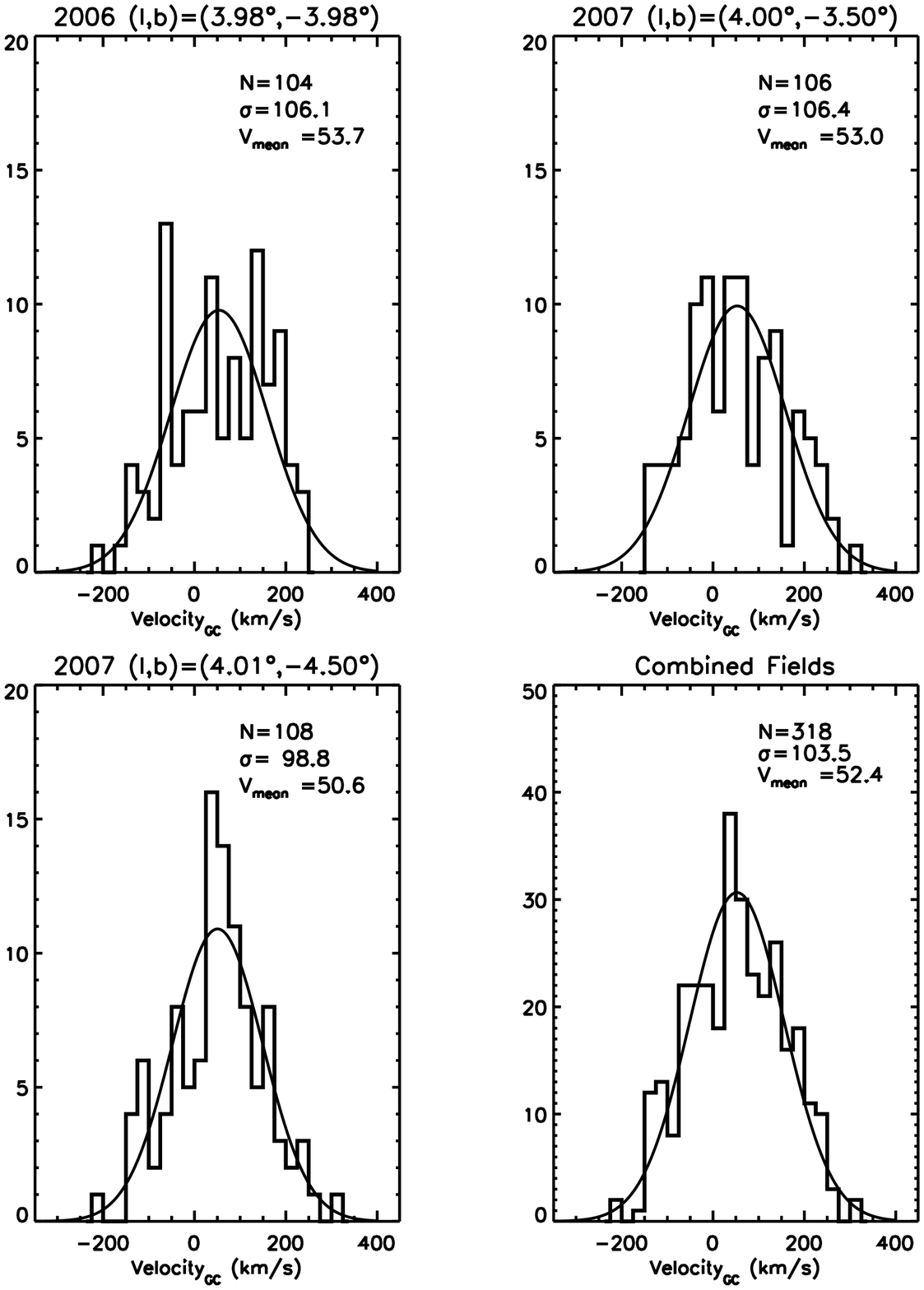}
\caption{Two candidate stream fields from 2006 were re-observed in 2007.  For each field location, the individual fields and statistics are shown in the first three panels, and the final summed field is shown in the final panel.  As can be seen, our initial detections of ''spikes'' turned out to be not significant, and emphasizes the importance of sample size when looking for cold stream features.  Bin size is 25 km~s$^{-1}$.}
\label{streams}
\end{figure*}

In 2006, hints of kinematic substructure in the velocity distributions of several fields were noted and reported in Rich et al. (2007b).  We now present the Galactocentric velocity distributions of all of our current bulge fields in figures \ref{all_hist} and \ref{all_hist_1}.  As can be seen, $\sim22$\% of the observed bulge fields show 2.5$\sigma$ deviations, as estimated by eye, from a Gaussian constructed from a calculation of mean velocity and dispersion in each field.  Indeed, the choice of bin width determines the amount of ''structure'' seen and at a bin width of 25 km~s$^{-1}$, we are perhaps overbinning our data.  The choice of bin width is driven by the $\sim10$ km~s$^{-1}$ dispersion expected for cold components like a dissolving cluster or dwarf galaxy.  Therefore, we believe it is still useful to display the data in this fashion in order to easily pick out features that may  justify follow-up observations.   Although the third (skew) and fourth (kurtosis) moments of the assumed Gaussian distributions calculated for each individual field was not significant, those fields with apparent deviation from a normal distribution warrant a closer inspection.  Possible substructure might arise from disrupted satellites or stars in unique orbit families but the relatively small number of stars observed in each field require subsequent observations to increase our sample size.   Our 2007 observations were aimed at investigating two of these fields to determine if the $~$2.5$\sigma$ deviations from the Gaussian in select 2006 fields (see Figure \ref{streams}) were real or merely statistical in nature. Two approaches were used to test these possible kinematic clumps.  For the field  ({\it l,b})$=$($-$8$^\circ$,$-$4$^\circ$), we shifted by half a degree in {\it b} above and below the original field, in order to detect a possible spike at the same velocity seen in the initial observation.  For the field ({\it l,b})$=$(4$^\circ$,$-$4$^\circ$), we obtained two separate exposures at the same (l,b), but selecting different stellar targets from the {\it 2MASS} catalog.  As can be seen in Figure \ref{streams}, the initial spikes turned out to be statistically not significant. However, we intend to re-observe other fields which exhibit similar structures.  Simulations by Reitzel et al. (2007) show that such deviations from a normal distribution are not unexpected in random draws, and preliminary examination suggested that roughly 16\% of the apparent substructures from our data set might be real.  A full statistical analysis between models and our data will be able to determine if we are seeing more ''spikes'' than are expected from random noise, but we will be unable to determine which one is real without gathering more data for each field. It is important to note that a dissolving system may have a range in abundance, so that the addition of abundance would not necessarily increase the significance of a detection.

\begin{figure*}[t]
\epsscale{1.2}
\plottwo{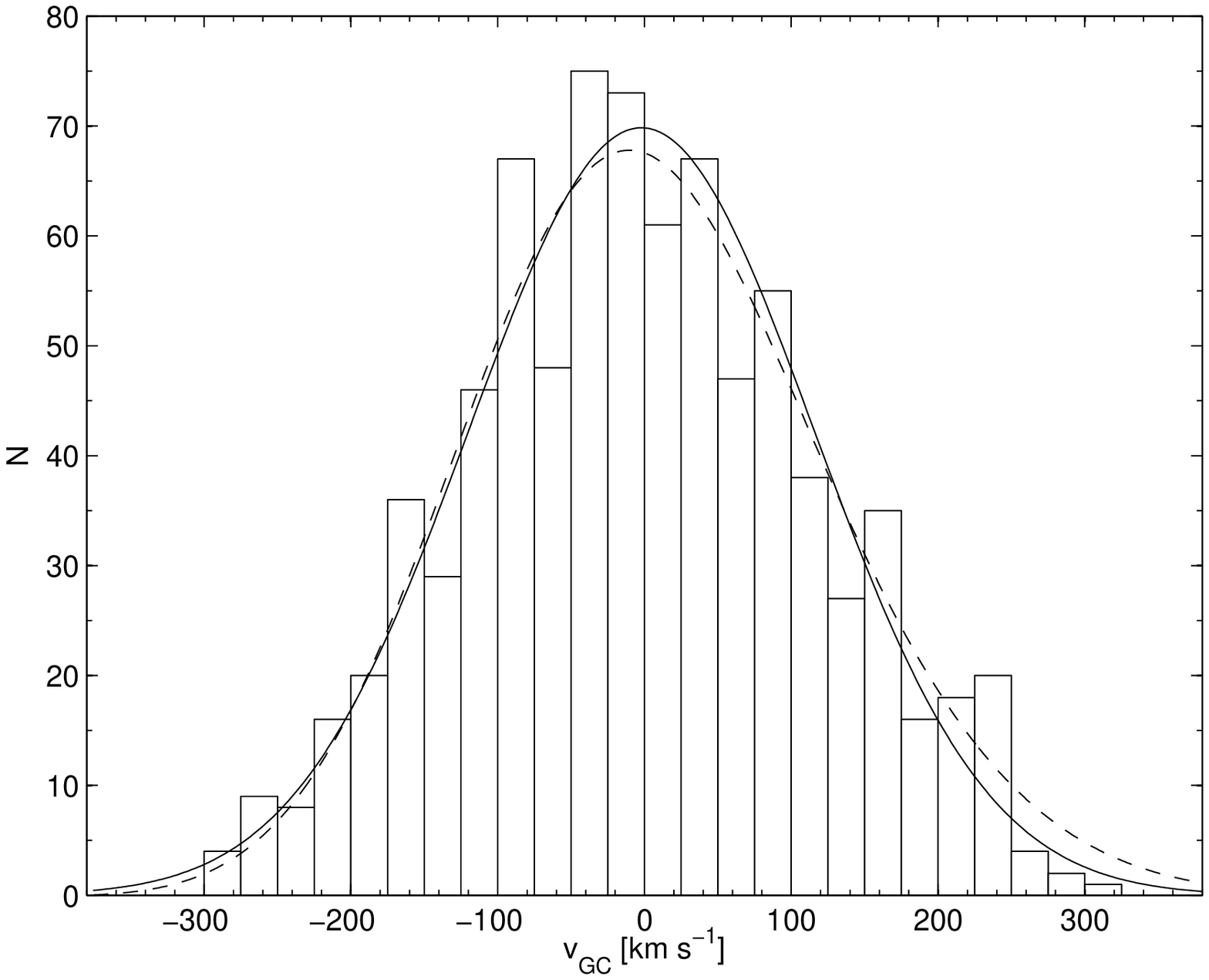}{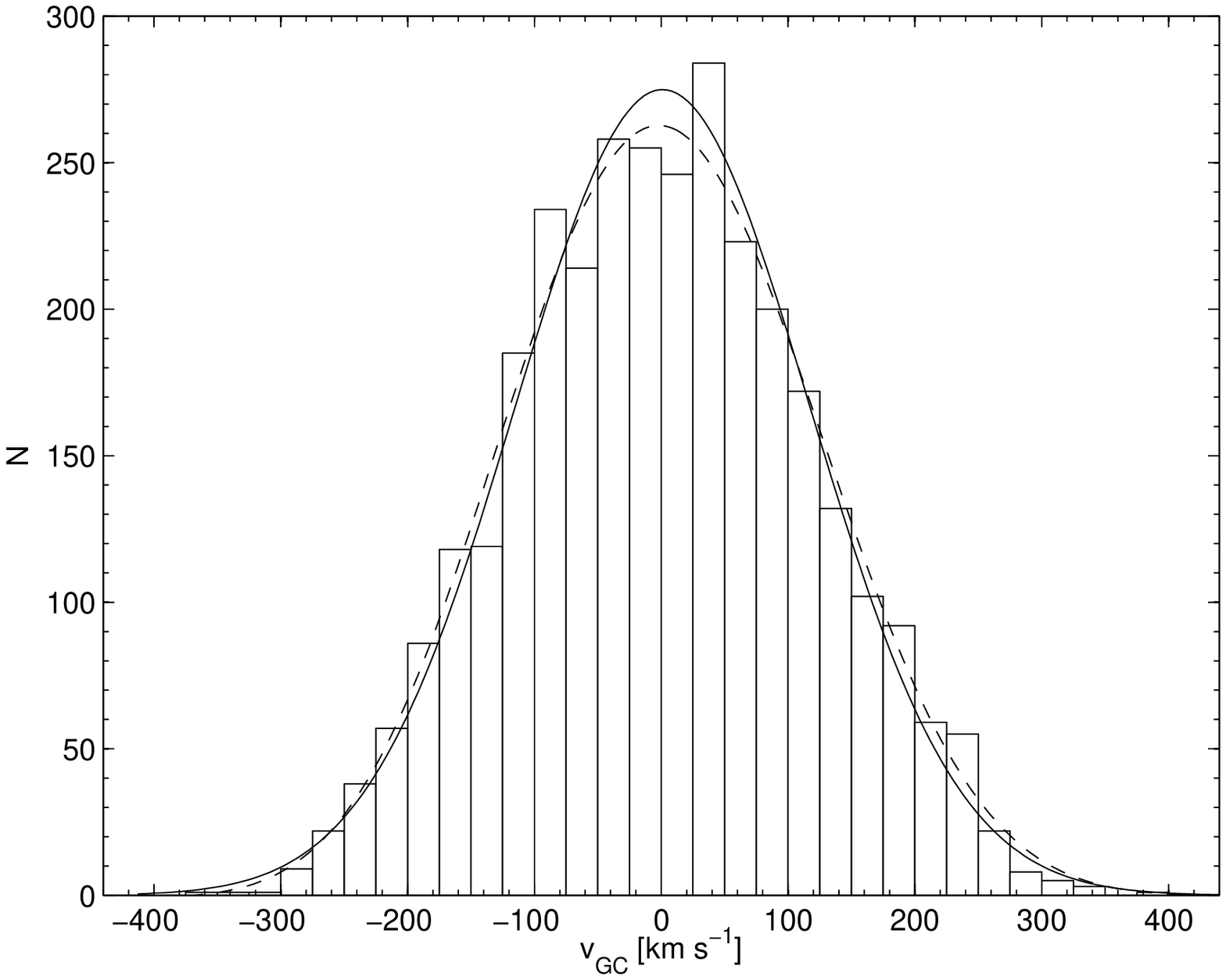}
\caption{Left: All minor axis fields summed, consisting of 822 stars.  Right: All bulge fields (both minor and major axes) summed, consisting of 3202 stars.  Both figures are in Galactocentric coordinates with a Gaussian with V$_{mean}$ and $\sigma$ derived from the IDL from a $\sigma$-clipping algorithm, 'meanclip', overlayed (solid line).  Also plotted is the fit of the Gauss-Hermite series (dotted line) as described in Section 6.  As can be seen, both curves are consistent with each other, and suggest our bulge sample consists of a homogeneous, normally distributed, stellar population.  Furthermore, measurements of the 3rd and 4th moments of the Gaussian distribution as well and the h\_3 and h\_4 parameters of the Gauss-Hermite series suggest no significant symmetric or assymetric deviations from a Gaussian distribution.}
\label{sum}
\end{figure*}

We now turn to the coadded total velocity distribution.  We have constructed the {\em minor axis} Galactocentric rest frame velocity distribution, giving V$_{mean}=-2 \pm 4$ km~s$^{-1}$ and $\sigma=117 \pm 3$ km~s$^{-1}$ for $\sim$820 stars, as well as having coadded the {\em entire} sample in Galactocentric coordinates, yielding V$_{mean}=1 \pm 2$ km~s$^{-1}$ and $\sigma=116 \pm 1$ km~s$^{-1}$ for $\sim$3200 stars (Figure \ref{sum}).  In both cases, the distribution appears Gaussian, with no apparent deviation from a normal distribution greater than $\approx$1.5$\sigma$, as estimated by eye.  For the entire coadded bulge sample, when fit by a $\chi^{2}$ minimization technique, we find a reduced $\chi^{2}=1.29$, with a mean of $0.6 \pm 2.4$ km~s$^{-1}$ and $\sigma=122 \pm 2$ km~s$^{-1}$, consistent with our calculations of the first and second moments of the distribution.  Furthermore, our distribution has a (negligible) skew of $0.05 \pm 0.04$ km~s$^{-1}$ and a kurtosis of $-0.36 \pm 0.09$ km~s$^{-1}$, suggesting a slightly platykurtic (flattened) distribution.  The negligible skew is in contrast to the measured distributions of SWC (1990; Figure 9) and Minniti (1992; Figure 1), which show strongly flattened or significantly skewed characteristics, indicating perhaps contaminated samples.  The lack of such characteristics in our distributions bolsters the argument that our samples are not contaminated by cold components (disk) or hot components (halo).

\begin{figure*}[t]
\epsscale{0.9}
\plotone{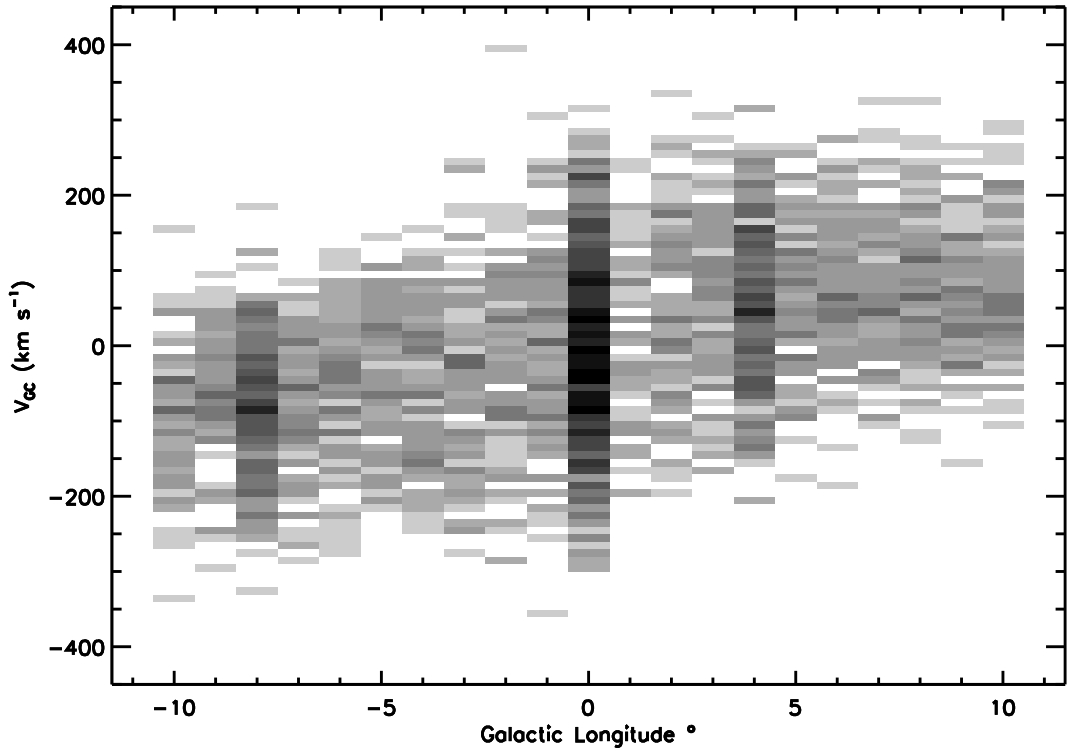}
\caption{Longitude-Velocity (l-v) plot for the entire bulge sample, smoothed to 1$^\circ$ in longitude and 10 km~s$^{-1}$ in galactocentric velocity.  The higher density of stars at l=0$^\circ$ is due to the inclusion of the minor axis fields, while the higher density at l=$-8$$^\circ$ and l=4$^\circ$ are due to our follow-up observations of those two fields in 2007 (see fig. \ref{streams}).  As can be seen, there is no evidence of a cold, disk component in our sample which would manifest itself as a linear trend.  We also do not see any evidence of a hot component that would indicate presence of a halo/spheroid population.  The roughly trapeoidal envelope of this distribution is reminiscent of that seen for the gas dynamics at R$<$2 kpc (Liszt \& Burton 1980).}
\label{l-v}
\end{figure*}

Furthermore, we have investigated characterizing our co-added samples using the Gauss-Hermite series as defined by van der Marel \& Franx (1993), which yield mean velocity, dispersion, as well as two other dimensionless parameters that measure the asymmetric (h\_3, analogous to skew) and symmetric (h\_4, analogous to kurtosis) deviations of a distribution from Gaussian (Figure \ref{sum}).  According to van der Marel \& Franx (1993), this method minimizes correlations between the parameters of the fit, and allows for a more sensitive measure of the deviation from a Gaussian line profile.    A least squares fitting to each summed distribution yields V$_{mean}=-2.6 \pm 5.4$ km~s$^{-1}$, $\sigma=122.2 \pm 5.3$ km~s$^{-1}$, h\_3$=0.038 \pm 0.031$, and h\_4$=0.002 \pm 0.031$ for the minor axis, and V$_{mean}$$=0.5 \pm 2.6$ km~s$^{-1}$, $\sigma=121.9 \pm 2.5$ km~s$^{-1}$, h\_3$=0.011 \pm 0.015$, and h\_4$=-0.014 \pm 0.015$ for the entire bulge sample.  These values are in excellent agreement with our more standard calculation of the moments of an assumed Gaussian distribution for the Galactic bulge. Figure \ref{l-v} shows our longitude-velocity plot for the major axis strip.  As can be seen, there is no evidence of a cold, disk component in our sample which would manifest itself as a linear trend.  There is no indication of a departure from a Gaussian single population in any field.  The overall trapezoidal appearance of the  outer envelope of the distribution is reminiscent of the gas dynamics observed by Liszt \& Burton (1980) within 2 kpc; the gas motions were modeled as arising in a tilted bar potential.  Our repeat tests with gaussian fitting, examination of the first four moments of the distribution, as well as the fits using the Gauss-Hermite series are all consistent with a purely Gaussian distribution, with no indication of contamination from any foreground stellar population. Furthermore, it is noteworthy that we discovered no stars with velocities greater than $\pm4\sigma$ in any of our fields.    

\subsection{ The Bulge on the (V$_{max}$/$\sigma$) vs. $\epsilon$ Plot}
\begin{figure*}[t]
\epsscale{0.9}
\plotone{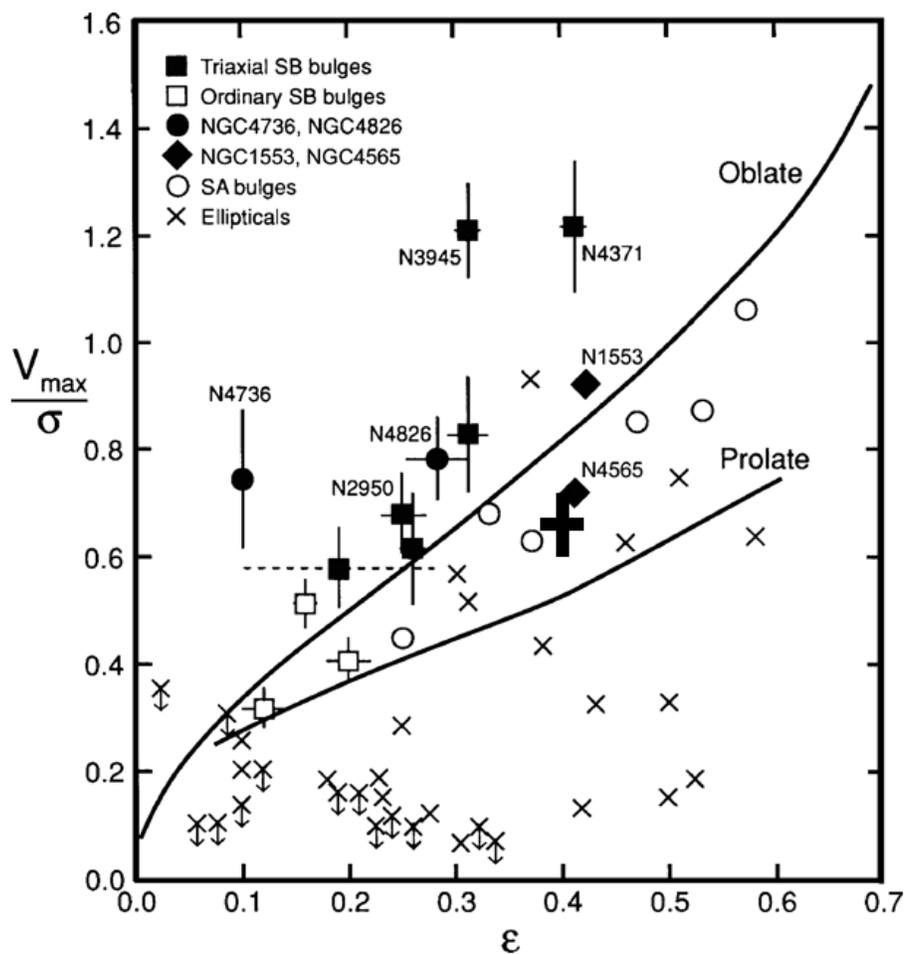}
\caption{(V$_{max}$/$\sigma$) plot from Kormendy \& Kennicutt (2004) with the Galactic bulge indicated (black cross).  The MW bulge lies under the oblate supported line and less rotationally supported than the pseudobulges (filled symbols) but is similar to classical bulges (open symbols).  NGC 4565 is an edge-on spiral, similar to the Milky Way, in that is exhibits a box-shaped bulge.}
\label{psuedo}
\end{figure*}

Armed with the velocity dispersion and a rotation speed, it is now possible to use the {\sl BRAVA} measurements to place the Galactic bulge on the Kormendy \& Kennicutt (2004) plot of (V$_{max}$/$\sigma$) vs. $\epsilon$ (Figure \ref{psuedo}).   One interprets the ratio (V$_{max}$/$\sigma$) as a ratio of ordered to random kinetic energy and $\epsilon$ as the apparent flattening of the bulge.  Placement of galaxies on this plot represent a method of interpreting the formation of bulge systems;  bulges formed via secular proceses (pseudo-bulges) lie on or above the oblate line and are more rotation dominated than classical bulges and ellipticals which are formed via mergers.  A very thorough discussion can be found in the review of Kormendy \& Kennicutt (2004), and shall not be repeated here.  We use the Weiland et al. (1994) minor-to-major axis ratio of $\approx0.6$ to estimate $\epsilon=1-$axial ratio$=0.4$.  We adopt $\sigma$$=$116 km~s$^{-1}$ for the velocity dispersion, and 75 km~s$^{-1}$ for the rotation speed, giving (V$_{max}$/$\sigma$)$=$0.64$\pm$0.5.  This places our Milky Way near the well-known edge-on spiral NGC 4565 with its peanut-shaped bulge, and near the oblate line, but below the most rapidly rotating bars and pseudobulges.  Kormendy \& Illingworth (1982) note that NGC 4565 not only has a peanut-shaped bulge, but exhibits cylindrical rotation (no decrease in rotation speed as a function of vertical distance above the plane), and therefore (V$_{max}$/$\sigma$) underestimates the dynamical importance of rotational kinetic energy compared to the ellipsoidal bulges.    While residing among the classical bulges in this plot, NGC 4565 and the Milky Way share the classic peanut bulge morphology while still lying well below the oblate line.  Kormendy \& Kennicutt (2004) note that a peanut-shaped bulge is sufficient to classify a galaxy has having a pseudobulge, so it may be no surprise that the Galactic bulge would lie beneath the oblate line should it indeed exhibit cylindrical rotation similar to NGC 4565.  While the stellar population of the bulge is old (see e.g. Ortolani et al. 1995) a pseudobulge can be old in the Kormendy \& Kennicutt (2004) definition.  The measurement of cylindrical rotation in the M giant population would strengthen substantially the analogy between the Milky Way's bulge and that of NGC 4565; indeed, we have observed this in the $-8^\circ$ latitude fields (Howard et al. 2008, in prep).  We defer consideration of the Milky Way's putative pseudobulge for this future work.

\section{Conclusions}

We present our methods of selecting M giant probes for the {\sl Bulge Radial Velocity Assay}.  Our sample as reported in this paper has nearly 3300 stars, and is defined clearly to cover the full range of abundance on the first red giant branch.  We report our rotation curve and velocity dispersion profiles, which agree more closely with the self consistent model of Zhao (1996) compared to N-body models in both rotation curve and dispersion.  

The reliability of our year-to-year measurements, along with our velocity precision, allow us to search for evidence of cold streams in the bulge resulting from recent merger events, analogous to the Sagittarius Dwarf spheroidal stream.  We find no evidence for hidden dynamically cold, or hot, subcomponents, either in the line of sight velocity distribution of the minor axis fields, or in the l-v plot.   Two candidate cold stream features are investigated in follow-up studies that increase the sample size by a factor of three.  The features were not confirmed, and illustrate the importance of sample size when searching for such features.

We find a departure from solid body rotation at a projected distance of $\approx$0.6 kpc for R$_{o}$=8 kpc, in agreement with the proper motion study of Clarkson et al. (private communication).  The break from solid body rotation occurs as the density of the bulge is dropping and may signify a transition from bar to keplerian dynamics.  The bulge appears to extend to $l=\pm 10^\circ$, beyond which a colder, more rapidly rotating, disk-like population appears to dominate.  However, it is noted that there is a lack of data between 10$^\circ$$<$$|$ {\it l} $|$$<$30$^\circ$, which prevents one from determining the relationship between our bulge sample, and the outer disk fields.  Our selection technique has also allowed us to determine that the the bulge population is normally distributed, with no significant deviation from Gaussian when considering the entire co-added sample of $\approx3200$ stars.

We have divided our sample by magnitude and color, and find no significant differences in dynamics.   We believe that our selection technique is robust and is yielding  a bulge/bar dominated sample.
The rotation curve permits us to place the Milky Way's bulge on the (V$_{max}$/$\sigma$) vs. $\epsilon$ plot; it is close to the peanut-shaped bulge of NGC 4565 in this diagram.  Further observations at the {\it b}$=-8^\circ$ latitude fields will allow us to determine if the Milky Way bulge rotates cylindrically, an indication of whether the Galactic bulge is a pseudobulge, or a more classically evolved bulge.

RMR and CDH acknowledge support from grant AST-0709479 from the National Science Foundation
and from grant GO-11159 from the Space Telescope Science Institute and NASA.   CDH acknowledges
support from a CTIO travel grant in support of doctoral thesis observations at CTIO.  The authors are grateful to Sylvie Beaulieau for providing her planetary nebula radial velocities for this work and Rich et al. (2007).  The authors also acknowledge useful comments by Will Clarkson, John Kormendy, and Dante Minniti.  The authors thank the support staff at CTIO.  This research has made use of the NASA/IPAC Extragalactic Database (NED) which is operated by the Jet Propulsion Laboratory, California Institute of Technology, under contract with the National Aeronautics and Space Administration.

\clearpage

\clearpage

\clearpage

\clearpage

\clearpage

\clearpage

\clearpage

\clearpage

\clearpage

\clearpage

\clearpage

\clearpage

\clearpage

\clearpage

\clearpage

\clearpage

\clearpage

\clearpage

\clearpage

\clearpage

\clearpage

\clearpage
\begin{deluxetable}{ccccccc}
\tabletypesize{\scriptsize}
\tablecaption{{\it BRAVA} Observation Details \label{log}}
\tablenotetext{\dagger}{Field P4 observed with the same configuration in 2005/2006}
\tablenotetext{\star}{3120s=1x300s,1x420s,1x600s,2x900s}
\tablewidth{0pt}
\tablehead{
\colhead{{\it BRAVA} field} & \colhead{Date of Obs.} & \colhead{Pointing Center}  & \colhead{Exposure} & \colhead{Grating} & \colhead{Central $\lambda$} & \colhead{Slit Mask} \\
\colhead{} & \colhead{(UT)} & \colhead{(J2000)}  & \colhead{} & \colhead{Angle} & \colhead{(\AA)} & \colhead{}}
\startdata
fa4                        & Jul 30 2005 & 18$^{h}$00$^{m}$25.02$^{s}$ -31$^\circ$58$^{\prime}$53$^{\prime\prime}$.1 & 1x300s,1x900s              & 18.206 & $\sim$7600\AA & -  \\
fe4                        & Jul 30 2005 & 17$^{h}$51$^{m}$43.92$^{s}$ -35$^\circ$19$^{\prime}$00$^{\prime\prime}$.1 & 3x300s                     & 18.206 & $\sim$7600\AA & -  \\
fm4                        & Jul 30 2005 & 18$^{h}$08$^{m}$17.54$^{s}$ -28$^\circ$21$^{\prime}$29$^{\prime\prime}$.5 & 3x300s                     & 18.206 & $\sim$7600\AA & -  \\
fo4                        & Jul 31 2005 & 18$^{h}$12$^{m}$31.04$^{s}$ -26$^\circ$36$^{\prime}$49$^{\prime\prime}$.4 & 2x420s                     & 18.206 & $\sim$7600\AA & -  \\
fp4-2005\tablenotemark{\dagger} & Jul 31 2005 & 18$^{h}$14$^{m}$40.35$^{s}$ -25$^\circ$42$^{\prime}$48$^{\prime\prime}$.7 & 3120s\tablenotemark{\star} & 18.206 & $\sim$7600\AA & -  \\
fMA                        & Aug 01 2005 & 17$^{h}$53$^{m}$27.22$^{s}$ -29$^\circ$58$^{\prime}$22$^{\prime\prime}$.4 & 2x900s,1x600s              & 18.206 & $\sim$7600\AA & -  \\
fMk                        & Aug 01 2005 & 17$^{h}$28$^{m}$31.48$^{s}$ -26$^\circ$31$^{\prime}$40$^{\prime\prime}$.2 & 2x900s                     & 18.206 & $\sim$7600\AA & -  \\
f1                         & May 16 2006 & 17$^{h}$36$^{m}$27.20$^{s}$ -39$^\circ$30$^{\prime}$23$^{\prime\prime}$.8 & 3x900s                     & 18.732 & $\sim$7800\AA & 200$\mu$m \\
f16                        & May 16 2006 & 18$^{h}$22$^{m}$52.16$^{s}$ -22$^\circ$10$^{\prime}$36$^{\prime\prime}$.1 & 3x900s                     & 18.732 & $\sim$7800\AA & 200$\mu$m  \\
f8                         & May 16 2006 & 17$^{h}$56$^{m}$53.48$^{s}$ -32$^\circ$41$^{\prime}$07$^{\prime\prime}$.7 & 3x900s                     & 18.732 & $\sim$7800\AA & 200$\mu$m  \\
f12                        & May 17 2006 & 18$^{h}$10$^{m}$19.68$^{s}$ -27$^\circ$28$^{\prime}$48$^{\prime\prime}$.1 & 2x600s,1x900s              & 18.732 & $\sim$7800\AA & 200$\mu$m  \\
f15                        & May 17 2006 & 18$^{h}$20$^{m}$47.74$^{s}$ -23$^\circ$03$^{\prime}$57$^{\prime\prime}$.9 & 2x300s,1x600s              & 18.732 & $\sim$7800\AA & 200$\mu$m  \\
f3                         & May 17 2006 & 17$^{h}$41$^{m}$55.61$^{s}$ -37$^\circ$50$^{\prime}$19$^{\prime\prime}$.3 & 3x600s,1x300s              & 18.732 & $\sim$7800\AA & 200$\mu$m  \\
 f5                         & May 17 2006 & 17$^{h}$47$^{m}$03.31$^{s}$ -36$^\circ$08$^{\prime}$10$^{\prime\prime}$.3 & 3x600s                     & 18.732 & $\sim$7800\AA & 200$\mu$m  \\
fswc                       & May 17 2006 & 18$^{h}$03$^{m}$44.62$^{s}$ -30$^\circ$01$^{\prime}$50$^{\prime\prime}$.0 & 3x600s                     & 18.732 & $\sim$7800\AA & 200$\mu$m  \\
f101                       & May 18 2006 & 17$^{h}$49$^{m}$43.50$^{s}$ -29$^\circ$56$^{\prime}$20$^{\prime\prime}$.9 & 2x600s,2x1200s             & 18.732 & $\sim$7800\AA & 200$\mu$m  \\
f13                        & May 18 2006 & 18$^{h}$16$^{m}$39.90$^{s}$ -24$^\circ$50$^{\prime}$11$^{\prime\prime}$.1 & 3x600s                     & 18.732 & $\sim$7800\AA & 200$\mu$m  \\
f14                        & May 18 2006 & 18$^{h}$18$^{m}$46.62$^{s}$ -23$^\circ$57$^{\prime}$49$^{\prime\prime}$.3 & 3x600s                     & 18.732 & $\sim$7800\AA & 200$\mu$m  \\
f4                         & May 18 2006 & 17$^{h}$44$^{m}$32.07$^{s}$ -36$^\circ$58$^{\prime}$46$^{\prime\prime}$.1 & 3x600s                     & 18.732 & $\sim$7800\AA & 200$\mu$m  \\
f7                         & May 18 2006 & 17$^{h}$54$^{m}$26.79$^{s}$ -33$^\circ$33$^{\prime}$12$^{\prime\prime}$.1 & 3x600s                     & 18.732 & $\sim$7800\AA & 200$\mu$m  \\
f11                        & May 19 2006 & 18$^{h}$06$^{m}$03.16$^{s}$ -29$^\circ$13$^{\prime}$30$^{\prime\prime}$.6 & 3x600s                     & 18.732 & $\sim$7800\AA & 200$\mu$m  \\
f2                         & May 19 2006 & 17$^{h}$39$^{m}$12.63$^{s}$ -38$^\circ$41$^{\prime}$39$^{\prime\prime}$.7 & 3x600s                     & 18.732 & $\sim$7800\AA & 200$\mu$m  \\
f6                         & May 19 2006 & 17$^{h}$52$^{m}$04.82$^{s}$ -34$^\circ$25$^{\prime}$09$^{\prime\prime}$.3 & 3x600s                     & 18.732 & $\sim$7800\AA & 200$\mu$m  \\
f9                         & May 19 2006 & 18$^{h}$01$^{m}$30.41$^{s}$ -30$^\circ$58$^{\prime}$26$^{\prime\prime}$.1 & 3x600s                     & 18.732 & $\sim$7800\AA & 200$\mu$m  \\
fp4-2006\tablenotemark{\dagger} & May 19 2006 & 18$^{h}$14$^{m}$40.35$^{s}$ -25$^\circ$42$^{\prime}$48$^{\prime\prime}$.7 & 3x600s                     & 18.732 & $\sim$7800\AA & 200$\mu$m  \\
f27                        & May 20 2006 & 18$^{h}$05$^{m}$38.05$^{s}$ -31$^\circ$26$^{\prime}$59$^{\prime\prime}$.1 & 3x600s                     & 18.732 & $\sim$7800\AA & 200$\mu$m  \\
f63                        & May 20 2006 & 18$^{h}$08$^{m}$42.54$^{s}$ -31$^\circ$51$^{\prime}$07$^{\prime\prime}$.0 & 3x600s                     & 18.732 & $\sim$7800\AA & 200$\mu$m  \\
f66                        & May 20 2006 & 17$^{h}$56$^{m}$46.67$^{s}$ -30$^\circ$43$^{\prime}$32$^{\prime\prime}$.1 & 4x600s                     & 18.732 & $\sim$7800\AA & 200$\mu$m  \\
f90                        & May 20 2006 & 17$^{h}$59$^{m}$33.84$^{s}$ -30$^\circ$43$^{\prime}$21$^{\prime\prime}$.1 & 3x600s                     & 18.732 & $\sim$7800\AA & 200$\mu$m  \\
f3030                      & Apr 07 2007 & 16$^{h}$23$^{m}$00.60$^{s}$ -55$^\circ$17$^{\prime}$29$^{\prime\prime}$.8 & 3x600s                     & 18.985 & $\sim$7900\AA & 200$\mu$m  \\ 
fa                         & Apr 25 2007 & 17$^{h}$41$^{m}$14.27$^{s}$ -38$^\circ$00$^{\prime}$06$^{\prime\prime}$.6 & 5x600s                     & 18.983 & $\sim$7900\AA & 200$\mu$m  \\
fb                         & Apr 25 2007 & 18$^{h}$08$^{m}$25.38$^{s}$ -27$^\circ$13$^{\prime}$55$^{\prime\prime}$.7 & 4x600s                     & 18.983 & $\sim$7900\AA & 200$\mu$m  \\
fc                         & Apr 25 2007 & 17$^{h}$41$^{m}$35.59$^{s}$ -37$^\circ$56$^{\prime}$12$^{\prime\prime}$.6 & 3x600s                     & 18.983 & $\sim$7900\AA & 200$\mu$m  \\
fd                         & Apr 25 2007 & 18$^{h}$12$^{m}$29.34$^{s}$ -27$^\circ$42$^{\prime}$40$^{\prime\prime}$.6 & 3x600s                     & 18.983 & $\sim$7900\AA & 200$\mu$m  \\

\enddata

\end{deluxetable}

\clearpage
\begin{deluxetable}{ccccc}
\tabletypesize{\scriptsize}
\tablecaption{Number of Reliable Velocity Measurements per Field \label{log1}}
\tablenotetext{\dagger}{Number of reliable velocities prior to color cuts. See section 5 for details}
\tablewidth{0pt}
\tablehead{
\colhead{{\it BRAVA} field} & \colhead{Pointing Center} & \colhead{Target Fibers in} & \colhead{Sky Fibers in} & \colhead{Reliable Velocities\tablenotemark{\dagger}} \\
\colhead{} & \colhead{(l$^\circ$,b$^\circ$)} & \colhead{Configuration} & \colhead{Configuration} & \colhead{}}
\startdata
fa4        &  -1.01, -4.29 & 110 & 20 & 109 \\
fe4        &  -4.81, -4.40 & 105 & 25 & 101 \\
fm4        &   2.99, -4.01 & 111 & 21 & 108 \\
fo4        &   4.98, -4.00 & 109 & 21 & 109 \\
fp4-2005   &   6.00, -4.00 & 106 & 20 & 101 \\
fMA        &  -0.01, -2.00 & 111 & 20 & 108 \\
fMk        &   0.01,  4.48 & 100 & 26 &  95 \\
f1         &  -9.98, -3.98 &  96 & 26 &  95 \\
f16        &  10.02, -3.99 & 107 & 21 & 106 \\
f8         &  -1.99, -3.99 & 109 & 19 & 109 \\
f12        &   3.98, -3.99 & 107 & 24 & 105 \\
f15        &   9.00, -3.99 & 105 & 22 & 101 \\
f3         &  -8.00, -4.00 & 103 & 25 & 102 \\
f5         &  -6.00, -3.99 & 106 & 22 & 105 \\
fswc       &   1.05, -3.96 &  74 & 20 &  72 \\
f101       &  -0.40, -1.28 & 100 & 31 &  97 \\
f13        &   6.99, -3.98 & 105 & 22 & 104 \\
f14        &   7.99, -4.00 & 108 & 22 & 107 \\
f4         &  -6.99, -4.00 & 104 & 25 & 103 \\
f7         &  -3.00, -3.98 & 112 & 19 & 109 \\
f11        &   2.00, -4.00 & 110 & 19 & 107 \\
f2         &  -9.01, -4.00 & 102 & 24 & 101 \\
f6         &  -4.00, -4.00 & 109 & 22 & 108 \\
f9         &  -0.01, -4.00 & 110 & 20 & 110 \\
fp4-2006   &   6.00, -4.00 & 109 & 21 & 108 \\
f27        &   0.00, -5.00 & 106 & 23 & 105 \\
f63        &  -0.04, -5.77 & 106 & 22 & 106 \\
f66        &  -0.30, -2.99 & 109 & 20 & 107 \\
f90        &   0.00, -3.51 & 112 & 20 & 112 \\
f3030      & -29.99, -3.96 & 100 & 20 &  96 \\ 
fa         &  -8.21, -3.97 & 108 & 17 & 100 \\
fb         &   4.00, -3.50 & 112 & 15 & 111 \\
fc         &  -8.11, -4.00 & 105 & 21 & 102 \\
fd         &   4.01, -4.52 & 111 & 17 & 109 \\

\enddata

\end{deluxetable}

\clearpage
\begin{deluxetable}{cccccc}
\tabletypesize{\scriptsize}
\tablecaption{2006/2007 Velocity Standards \label{standards}}
\tablewidth{0pt}
\tablehead{
\colhead{Indentifier} & \colhead{RA} & \colhead{DEC}  & \colhead{Spectral Type} & \colhead{Velocity} & \colhead{Year Observed}\\
\colhead{} & \colhead{(J2000)} & \colhead{(J2000)} & \colhead{} & \colhead{(km~s$^{-1}$)} & \colhead{}}
\startdata
HD 203638 & 21$^{h}$24$^{m}$09.593$^{s}$ & -20$^\circ$51$^{\prime}$06$^{\prime\prime}$.73 & K0III   &  22.0 & 2006 \\
HD 177017 & 19$^{h}$03$^{m}$43.756$^{s}$ & -22$^\circ$42$^{\prime}$43$^{\prime\prime}$.00 & M7III   &  42   & 2006 \\
HD 207076 & 21$^{h}$46$^{m}$31.849$^{s}$ & -02$^\circ$12$^{\prime}$45$^{\prime\prime}$.92 & M8IIIv  & -37.2 & 2006 \\
HD 218541 & 23$^{h}$09$^{m}$05.561$^{s}$ & -30$^\circ$08$^{\prime}$02$^{\prime\prime}$.15 & M6III   &  30.0 & 2006 \\
HD 134140 & 15$^{h}$08$^{m}$57.514$^{s}$ & -26$^\circ$29$^{\prime}$50$^{\prime\prime}$.18 & M1III   &  34.8 & 2007 \\
HD 146051 & 16$^{h}$14$^{m}$20.739$^{s}$ & -03$^\circ$41$^{\prime}$39$^{\prime\prime}$.56 & M0.5III & -19.9 & 2007 \\

\enddata

\end{deluxetable}

\clearpage
\begin{deluxetable}{ccccccccc}
\tabletypesize{\scriptsize}
\tablecaption{{\it BRAVA} Rotation and Dispersion Results \label{tbl-1}}
\tablenotetext{\dagger}{Summed result of previous 3 fields.  For plotting, these values are used for this ({\it l,b}).  See also Fig. \ref{streams}}
\tablenotetext{\star}{Number of velocities after color cut.  This value is used to calculate individual field statistics and err($<$V$>$) and err($\sigma$). }
\tablenotetext{\star\star}{Number of reliable velocities before color cut.  This value is the same as reported in Table \ref{log1}}
\tablewidth{0pt}
\tablehead{
\colhead{{\it BRAVA} field} & \colhead{l,b} & \colhead{$<$V$_{HC}>$} & \colhead{$<$V$_{GC}>$} & \colhead{err($<$V$>$)} & \colhead{$\sigma$} & \colhead{err($\sigma$)} & \colhead{N\tablenotemark{\star}(N\tablenotemark{\star\star})} \\
\colhead{} & \colhead{(degrees)} & \colhead{(km~s$^{-1}$)} & \colhead{(km~s$^{-1}$)} & \colhead{(km~s$^{-1}$)} & \colhead{(km~s$^{-1}$)} & \colhead{(km~s$^{-1}$)} & \colhead{}}
\startdata
f3030       &-30.00, -3.98 & -53.3 & -161.7 &   6.1 &  51.3 &  4.3 &   70(96) \\
f1          & -9.97, -3.99 & -66.9 &  -98.6 &   9.3 &  84.1 &  6.6 &   81(95) \\
f2          & -9.01, -4.01 & -35.5 &  -63.4 &   8.3 &  79.7 &  5.8 &  93(101) \\
fa          & -8.20, -3.98 & -68.5 &  -93.1 &   9.2 &  86.8 &  6.5 &  89(100) \\
fc          & -8.09, -4.00 & -64.0 &  -88.1 &   8.7 &  83.2 &  6.2 &  91(102) \\
f3          & -7.97, -4.00 & -45.9 &  -69.6 &   8.4 &  81.1 &  5.9 &  94(102) \\
f3-all\tablenotemark{\dagger}      & -8.09, -3.99 & -59.4 &  -83.6 &   5.1 &  84.2 &  3.6 & 272(300) \\
f4          & -6.97, -3.99 & -62.6 &  -82.2 &   8.5 &  85.7 &  6.0 & 102(103) \\
f5          & -6.00, -3.99 & -44.6 &  -60.4 &   8.9 &  88.6 &  6.3 & 100(105) \\
fe4         & -4.81, -4.38 & -25.8 &  -36.8 &   9.0 &  90.3 &  6.4 & 101(101) \\
f6          & -3.99, -3.98 & -44.7 &  -52.3 &   9.3 &  96.5 &  6.6 & 108(108) \\
f7          & -3.00, -3.99 & -46.7 &  -50.4 &  10.7 & 111.9 &  7.6 & 109(109) \\
f8          & -2.00, -3.98 & -23.2 &  -22.8 &  11.0 & 112.8 &  7.7 & 106(109) \\
fa4         & -1.00, -4.28 &  -9.9 &   -5.5 &  11.4 & 119.1 &  8.1 & 109(109) \\
f101        & -0.39, -1.29 &  -1.5 &    5.8 &  15.0 & 135.4 & 10.6 &   82(97) \\
f66         & -0.30, -2.99 & -27.8 &  -20.4 &  11.9 & 121.6 &  8.4 & 105(107) \\
f63         & -0.02, -5.78 & -11.4 &   -3.2 &  10.5 & 108.1 &  7.4 & 106(106) \\
f9          & -0.02, -3.99 & -14.9 &   -6.5 &  10.8 & 112.8 &  7.6 & 110(110) \\
fMA         & -0.01, -2.00 & -16.8 &   -8.1 &  12.1 & 125.8 &  8.6 & 102(108) \\
f27         & -0.01, -5.00 & -11.9 &   -3.6 &   9.7 &  99.0 &  6.8 & 105(105) \\
f90         &  0.00, -3.51 &   5.4 &   13.9 &  11.8 & 124.6 &  8.4 & 111(112) \\
fMk         &  0.00,  4.50 &  -1.8 &    7.7 &  11.6 & 112.9 &  8.2 &   95(95) \\
fswc        &  1.05, -3.96 &  -4.7 &    8.0 &  14.3 & 111.5 & 10.1 &   61(72) \\
f11         &  2.00, -4.00 &  30.5 &   47.0 &  11.3 & 116.3 &  8.0 & 106(107) \\
fm4         &  3.01, -4.01 &  37.1 &   57.8 &  10.1 & 104.7 &  7.1 & 108(108) \\
f12         &  3.99, -3.99 &  29.2 &   53.7 &  10.4 & 106.1 &  7.4 & 104(105) \\
fb          &  4.00, -3.50 &  28.3 &   53.0 &  10.3 & 106.4 &  7.3 & 106(111) \\
fd          &  4.01, -4.50 &  26.0 &   50.6 &   9.5 &  98.8 &  6.7 & 108(109) \\
f12-all\tablenotemark{\dagger}     &  4.00, -4.00 &  27.8 &   52.4 &   5.8 & 103.5 &  4.1 & 318(325) \\  
fo4         &  4.99, -3.99 &  31.8 &   60.4 &   9.1 &  94.8 &  6.4 & 109(109) \\
fP4-2006    &  6.01, -3.99 &  49.3 &   82.0 &   8.5 &  88.2 &  6.0 & 108(108) \\
fP4-2005    &  6.02, -4.00 &  44.9 &   77.7 &   9.1 &  91.6 &  6.4 & 101(101) \\
f13         &  6.99, -3.98 &  35.7 &   72.3 &   9.5 &  91.7 &  6.7 &  93(104) \\
f14         &  7.99, -3.99 &  44.0 &   84.6 &   8.5 &  88.2 &  6.0 & 107(107) \\
f15         &  9.01, -3.98 &  22.2 &   66.8 &   8.1 &  77.8 &  5.7 &  92(101) \\
f16         & 10.00, -3.99 &  35.7 &   84.2 &   8.8 &  87.0 &  6.2 &  97(106) \\

\enddata

\end{deluxetable}

\end{document}